\documentclass[reprint,amsmath,amssymb,aps]{revtex4-2}
\usepackage{textcomp}
\usepackage[utf8]{inputenc}
\usepackage{color}
\usepackage{cprotect}
\usepackage{float}
\usepackage{bm}
\usepackage{amsmath}
\usepackage{graphicx}
\usepackage[bookmarks=true,bookmarksnumbered=false,bookmarksopen=false,
 breaklinks=false,pdfborder={0 0 0},pdfborderstyle={},backref=false,colorlinks=true]
 {hyperref}
\hypersetup{pdftitle={Characterizing Neural Manifolds’ Properties and Curvatures using Normalizing Flows},
 pdfauthor={Peter Bouss, Sandra Nestler, Kirsten Fischer, Claudia Merger, Alexandre Rene and Moritz Helias},
 linkcolor=blue, citecolor=blue, urlcolor=black, filecolor=blue}

\makeatletter

\newcommand*\LyXZeroWidthSpace{\hspace{0pt}}
\providecommand{\tabularnewline}{\\}

\usepackage{dcolumn}
\usepackage{bm}

\makeatother

\begin{document}

\title{Characterizing Neural Manifolds’ Properties and Curvatures\\ using Normalizing Flows}

\author{Peter Bouss\textsuperscript{1,2}}
\email{p.bouss@fz-juelich.de}
\thanks{These authors contributed equally to this work.}

\author{Sandra Nestler\textsuperscript{1,2,3}}
\email{p.bouss@fz-juelich.de}
\thanks{These authors contributed equally to this work.}

\author{Kirsten Fischer\textsuperscript{1,2}}

\author{Claudia Merger\textsuperscript{1,4,5}}
\author{Alexandre Ren\'{e}\textsuperscript{1,6,7}}
\author{Moritz Helias\textsuperscript{1,4}}

\affiliation{\textsuperscript{1}Institute for Advanced Simulation (IAS-6), Jülich Research Centre, Jülich, Germany}
\affiliation{\textsuperscript{2}RWTH Aachen University, Aachen, Germany}
\affiliation{\textsuperscript{3}Rappaport Faculty of Medicine and Network Biology Research Laboratory, Technion - Israel Institute of Technology, Haifa, Israel}
\affiliation{\textsuperscript{4}Department of Physics, Faculty 1, RWTH Aachen University, Aachen, Germany}
\affiliation{\textsuperscript{5}International School of Advanced Studies (SISSA), Trieste, Italy}
\affiliation{\textsuperscript{6}Department of Physics, University of Ottawa, Ottawa, Canada}
\affiliation{\textsuperscript{7}Department of Computer Science, RWTH Aachen University, Aachen, Germany}

\date{\today}

\begin{abstract}
Neuronal activity is found to lie on low-dimensional manifolds embedded
within the high-dimensional neuron space. Variants of principal component
analysis are frequently employed to assess these manifolds. These
methods are, however, limited by assuming a Gaussian data distribution
and a flat manifold. In this study, we introduce a method designed
to satisfy three core objectives: (1) extract coordinated activity
across neurons, described either statistically as correlations or
geometrically as manifolds; (2) identify a small number of latent
variables capturing these structures; and (3) offer an analytical
and interpretable framework characterizing statistical properties
by a characteristic function and describing manifold geometry through
a collection of charts.

To this end, we employ Normalizing Flows (NFs), which learn an underlying
probability distribution of data by an invertible mapping between
data and latent space. Their simplicity and ability to compute exact
likelihoods distinguish them from other generative networks. We adjust
the NF’s training objective to distinguish between relevant (in manifold)
and noise dimensions (out of manifold). Additionally, we find that
different behavioral states align with the components of the latent
Gaussian mixture model, enabling their treatment as distinct curved
manifolds. Subsequently, we approximate the network for each mixture
component with a quadratic mapping, allowing us to characterize both
neural manifold curvature and non-Gaussian correlations among recording
channels.

Applying the method to recordings in macaque visual cortex, we demonstrate
that state-dependent manifolds are curved and exhibit complex statistical
dependencies. Our approach thus enables an expressive description
of neural population activity, uncovering non-linear interactions
among groups of neurons. 
\end{abstract}
\maketitle
\maketitle

\section{Introduction}

\begin{figure*}
\includegraphics{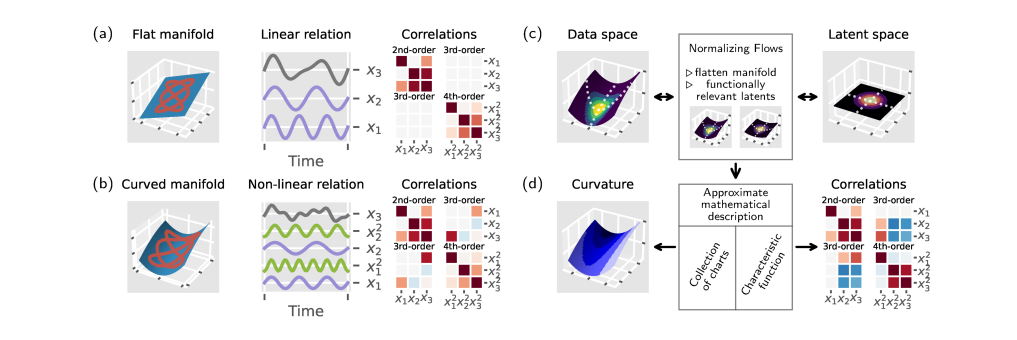}\caption{\protect\label{fig:graphical-abstract}\textbf{Analysis pipeline for
curved neural manifolds parametrized in terms of meaningful latent
variables. (a) }A conceptual signal trajectory $x_{i}$ (red) embedded
into a three-dimensional state space (left panel) but lying on a flat
manifold (blue). Each axis represents one data channel. The middle
panel displays the time evolution of each channel. The first two signals
$x_{1}$ and $x_{2}$ (in purple) are independent oscillatory signals,
while $x_{3}$ (in gray) is a weighted sum of $x_{1}$ and $x_{2}$.
The relation is thus linear. In the right panel, correlations of 2\protect\textsuperscript{nd},
3\protect\textsuperscript{rd} and 4\protect\textsuperscript{th}~order
are shown for combinations of $x_{i}$ or $x_{i}^{2}$ and $x_{j}$
or $x_{j}^{2}$. The 3\protect\textsuperscript{rd}~order correlations
vanish. The 4\protect\textsuperscript{th}~order correlations result
from the 2\protect\textsuperscript{nd}~order correlations only.
\textbf{(b)} Left panel similar to a): a trajectory (red) of a signal
$x_{i}$ is shown that lies on a manifold, which is now curved (blue).
The construction can be seen in the middle panel. The signal $x_{3}$
(gray) depends here on $x_{1}$ and $x_{2}$ (purple) but also on
$x_{1}^{2}$ and $x_{2}^{2}$ (green). The manifold is curved due
to this non-linear relation. In contrast to a), the right panel shows
non-zero 3\protect\textsuperscript{rd}~order correlations, which
form a signature of the manifold’s curvature. \textbf{(c) }Left panel
shows a probability density lying on a curved manifold. The Normalizing
Flow (middle panel) learns to map both the distribution and its underlying
manifold to a latent distribution on a flat manifold (right panel).
In this deep neural network, every layer implements an invertible
transformation. The latent distribution, described as a mixture of
Gaussians, lies in a space whose axes (white dashed lines) correspond
to the learned curved axes as shown in the original distribution (left
panel); these are functionally relevant latent axes. \textbf{(d) }The
trained Normalizing Flow is approximated by a quadratic expression,
which can be used to create a collection of charts and a characteristic
function (middle panel). The charts can be used to compute the curvature
of the manifold (left panel). The characteristic function yields correlations
of different orders (right panel).}
\end{figure*}

Understanding the neural code used by the brain to represent information
is a central quest in neuroscience. \citep{Georgopoulos86,Seung93_10749,Olshausen1996emergence,Sanger96,Gerstner97,Knill04_712,Pouget06_1432,Doya07,DiCarlo12_415,Saxena19_103}.
The question of representation encompasses two distinct aspects. The
first concerns the relations between the activities of multiple cells.
These relations can either be described \emph{statistically}, for
example in terms of correlations among the activities of multiple
cells, or \emph{geometrically}, in terms of the subspaces spanned
by the joint activity of the entire neuronal population, also termed
neural manifolds.

The second aspect of the neural code concerns the relation of the
neuronal activity to informative lower-dimensional latent variables~\citep{Jazayeri2021interpreting,Recanatesi2021predictive}.
These latent variables can for instance represent movement parameters~\citep{Churchland12_51,Kaufman2014cortical,Sussillo15_1025,Kao2021optimal},
encode sensory stimuli \citep{Mante13_78,Stringer19_361} or decision-making
processes~\citep{Kim2021inferring,Steinemann2024direct,Langdon2025latent},
or reflect other continuous internal states \citep{Pandarinath18_805,Whiteway2019quest}.
Inferring low-dimensional latent structure from high-dimensional neuronal
activity is thus a key objective of computational neuroscience. Deciphering
the neural code amounts to finding both, a description of the statistical
or geometrical relations among the activities within a neural population,
as well as identifying how latent causes are linked to this observed
activity.

A recent line of research studies trajectories of neural activity
in the space spanned by the activity of all neurons \citep{Gallego17_978,Gallego18_1,Gallego20_260,Altan2021estimating,Altan2023low,Langdon2023unifying,Safaie2023preserved,Fortunato2024nonlinear,MoralesGregorio2024neural}
and has found that these trajectories indeed do not fill the entire
space, but rather lie on manifolds with a lower intrinsic dimensionality
than that of the embedding space \citep{Gao17_4262,Altan2021estimating,Jazayeri2021interpreting,Langdon2023unifying}.
Such manifolds imply that neurons are not activated independently
of each other, but must rather obey intricate dependencies. Gaining
insight into the composition, structure, and properties of these manifolds
is central to understand how the brain represents and processes information.
A number of studies on neural manifolds have been carried out for
different brain areas, such as motor cortex \citep{Churchland12_51,Kaufman2014cortical},
prefrontal cortex \citep{Druckmann_12,Mante13_78,Kobak2016demixed},
hippocampus \citep{Gardner22_123,Schneider2023learnable}, visual
cortex \citep{DiCarlo12_415,Stringer19_361}, and basal ganglia \citep{Gallego17_978,Gallego18_1}. 

Principal Component Analysis (PCA) \citep{Pearson1901,Hotelling1933analysis}
is the most prominent method used in these studies \citep{Mitchell2023neural},
along with its many variants \citep{Bishop1998bayesian,Zou2006sparse,Churchland12_51,Kobak2016demixed}
including tensor-based methods \citep{Williams2018unsupervised,Pellegrino2024dimensionality}.

PCA and related methods have been key in identifying low-dimensional
structure, but their assumption of Gaussianity of the data distribution
and the restriction to linear operations limit their ability to describe
curved manifolds or complex dependencies in the neural activity. As
shown in Fig.~\ref{fig:graphical-abstract}a), a linear, Gaussian
description suffices when the data lie on a flat manifold -- for
example, when $x_{3}$ is a linear combination of $x_{1}$ and $x_{2}$.
In this case, 3\textsuperscript{rd}~order correlations vanish and
4\textsuperscript{th}~order correlations result only from 2\textsuperscript{nd}~order
statistics.

These assumptions, however, rarely hold for neural recordings from
animals \citep{Duncker2021dynamics,Abbaspourazad2024dynamical,Fortunato2024nonlinear}:
these often exhibit non-Gaussian statistics of the neural activity
and curvature in the neural manifold -- characteristics necessitating
non-linear methods for effective description and analysis. In Fig.~\ref{fig:graphical-abstract}b),
we show an example of a curved manifold: here the variable $x_{3}$
is quadratically dependent on the other variables, leading to a curvature
of the data manifold in state space, as well as non-trivial higher-order
correlations between variables.

Various non-linear methods, including variational autoencoders \citep{Kingma15,Han2019variational,Zhou2020learning,Schimel2021ilqr,Ahmed2022examining},
have been proposed to characterize data on curved manifolds~\citep{Tenenbaum00_2319,Roweis2000nonlinear,Belkin2003laplacian,VanderMaaten2008visualizing,Mcinnes2018umap,Lindenbaum2018geometry,Schneider2023learnable},
many relying on generative neural networks~\citep{Gemici2016normalizing,Pandarinath18_805,Rey2019diffusion,Brehmer2020flows,Kim2020softflow,Caterini2021rectangular,Horvat2021denoising,Koehler2021variational,Connor2021variational,Cunningham2022principal,Cramer2022nonlinear,Postels2022maniflow,Horvat2022intrinsic,Loaiza2022diagnosing,Horvat2023density,Flouris2024canonical,Alberti2024manifold}.
While these methods describe certain aspects of curved manifolds well,
many of them share the shortcomings that they are difficult to interpret
in terms of describing the animal's behavior with latent variables
or they lack an explicit description of the curved underlying manifold
geometry. Our goal with this work is to address these shortcomings:
we introduce a method that captures both the geometric structure and
the resulting statistical dependencies within the data, while also
offering a low-dimensional space of latent variables. Both geometry
and statistics are extracted from the same analytical approximation
of the mapping between latent and data space.

Concretely, in this work we introduce a method which simultaneously
(1) learns a low-dimensional latent representation, (2) characterizes
the data manifold’s geometry, and (3) derives an interpretable description
of the manifold’s curvature and yields high-order statistics of the
data. To the best of our knowledge, this is the first method to achieve
this combination. The proposed method opens the door to a richer understanding
of how neuronal activity is structured, beyond what current generative
or dimensionality reduction techniques can provide.

To achieve this, we train invertible neuronal networks (also known
as a Normalizing Flows (NFs))~\citep{DInh15_1410,Dinh2016density,Kingma2018glow}
to learn a bijective mapping from the data space to a latent space;
the training is such that this latent space is flat and the latent
variables follow a simple distribution. Although there have been proposals
to learn manifolds with Normalizing Flows~\citep{Brehmer2020flows,Kim2020softflow,Caterini2021rectangular,Horvat2021denoising,Cramer2022nonlinear,Horvat2022intrinsic,Loaiza2022diagnosing,Postels2022maniflow,Horvat2023density,Flouris2024canonical},
these do not address the three objectives that we aim for in this
study. With our approach, we propose three key extensions to regular
NFs: the first two are concerned with the design of the loss function
and the latent space, and the third improves the interpretability
of the learned manifold.

First, we enforce a low-dimensional structure in the latent space
by a modified loss function \citep{Bekasov2020ordering}, ensuring
that the most relevant directions in latent space correspond to meaningful
neural population dynamics. Mapping backwards from the latent to the
data space, we thus obtain a curved, low-dimensional manifold in the
data space that captures the most prominent features of the data.
This is sketched in Fig.~\ref{fig:graphical-abstract}c), where the
axes of the latent space are shown as white dotted lines; when they
are mapped backwards through the network, they serve as informative,
curved axes describing the original data distribution.

Second, we replace the standard Gaussian latent distribution with
a mixture of Gaussians. This allows the network to represent multimodal
neural data -- an essential feature for capturing state-dependent
activity.

Third, to enhance interpretability, we approximate the learned network
mapping with a quadratic function. We then obtain the correlations
between neuronal activations by extracting an approximate characteristic
function from the network mapping (see Fig.~\ref{fig:graphical-abstract}d),
right). This approximation also allows us to obtain a set of differential
geometry charts which in turn yield the Riemannian curvature tensor
of the neural manifold (see Fig.~\ref{fig:graphical-abstract}d),
left). 

To showcase the utility of our approach, we apply it to neural recordings
from the macaque visual cortex \citep{Chen22_77}. In this data, we
find different components, where some of them clearly align to the
behavioral states. The other components show clear non-Gaussian higher-order
correlations, and geometrically, these components lie on manifolds
that are saddle-point-like. These components are not so closely tied
to one of the behavioral states. The components which are stronger
aligned to the behavioral states show deviations from Gaussianity
to a lesser degree and lie on less curved manifolds.

The remainder of the work is organized as follows:

Sec.~\ref{subsec:data-overview} introduces the macaque visual cortex
dataset and highlights the deviations from Gaussianity, with a focus
on higher-order correlations. In Sec.~\ref{sec:normalizing-flows},
we provide a concise overview into Normalizing Flows. Sec.~\ref{sec:reconstruction-error}
outlines our additional reconstruction error loss function and illustrates,
through a synthetic example, how it reveals a curved manifold in data.
We then introduce a multimodal latent space in Sec.~\ref{sec:multimodality},
explaining how this enables the analysis of datasets with different
behavioral states. Sec.~\ref{subsec:number-of-components} details
the selection of an appropriate multimodal latent space for our macaque
visual cortex data.

We evaluate the alignments of the components that our model gives
to the ground-truth behavioral states of the data in Sec.~\ref{subsec:label-matching}.
Next, in Sec.~\ref{subsec:quadratic-approximation}, we introduce
a quadratic function as an approximation of the inverse map of the
network, forming the basis for further analytical insight. Sec.~\ref{subsec:cumulants}
leverages this approximation to compute the moment- and cumulant-generating
functions for each of the components and analyzes higher-order statistical
structure for each of them. The geometric properties of the inferred
manifolds are explored in Sec.~\ref{subsec:curvature}, where we
identify saddle-point-like curvatures.

Lastly, Sec.~\ref{sec:Discussion} situates our approach within the
broader literature, compares it to related models, discusses limitations,
and outlines promising directions for future work.

\section{Correlation and geometry effects in the neural code\protect\label{subsec:data-overview}}

\begin{figure*}
\includegraphics{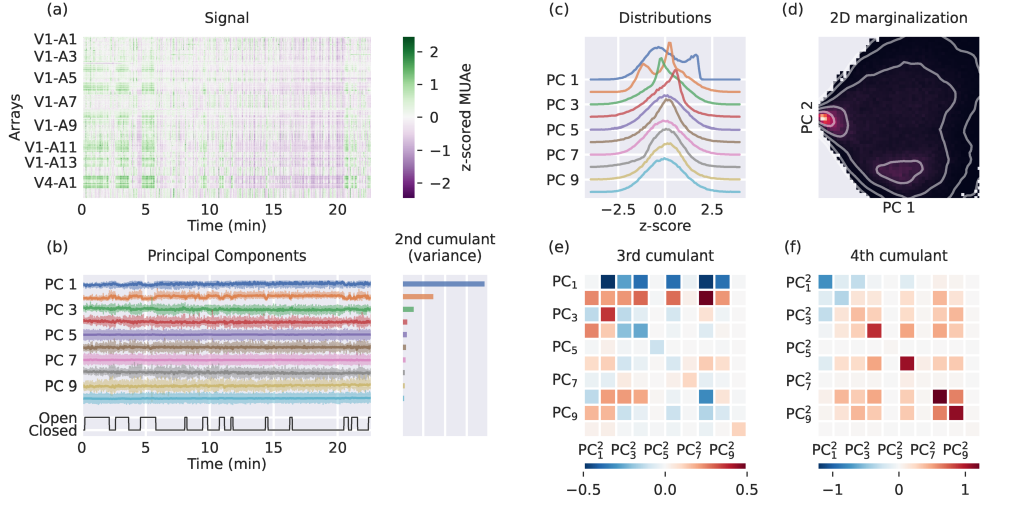}\caption{\protect\label{fig:data-overview}\textbf{Neuronal data exhibit higher-order
correlations. (a) }The original signal from one recording session
(L\_RS\_250717), sampled at 100~Hz over 22.7~minutes. The signal
is a multi-unit activity envelope (MUAe) containing 774 channels from
16 recording electrode arrays (14 in V1 and 2 in V4). Each channel
is independently $z$-scored (zero mean, unit variance) to ensure
consistent scaling across neurons. \textbf{(b) }Time evolution of
the first ten PCA variables (PCs) of the $z$-scored data. Transparent
curves represent raw PCs, while solid curves show temporal smoothing
using a Gaussian kernel with 1~s width. The labels of the behavioral
state (eyes open or eyes closed) are shown below. The rapid decay
of the eigenvalues $\sigma_{i}^{2}$ of the covariance matrix (2nd-order
cumulants; on the right), corresponding to the shown PCs, is an indicator
that the neural data lie on a lower-dimensional manifold. \textbf{(c)
}Marginal distribution of the first ten PCs. While some appear Gaussian,
others clearly deviate from Gaussianity. \textbf{(d)} Two-dimensional
marginalization along the first and second PCs. \textbf{(e-f)} 3\protect\textsuperscript{rd}
and 4\protect\textsuperscript{th}~order cumulant. For both cases,
the signals along the PCs are first divided by their standard deviation
$\sigma_{i}$ to normalize for the amplitude. Thus, panel (e) portrays
$x_{i}/\sigma_{i}$ against $x_{j}^{2}/\sigma_{j}^{2}$ and panel
(f) $x_{i}^{2}/\sigma_{i}^{2}$ against $x_{j}^{2}/\sigma_{j}^{2}$.
Color bars indicate cumulant magnitudes.}
\end{figure*}

To investigate the representation of information in cortical neuronal
networks, we use a set of electrophysiological recordings with multiple
hundreds of channels, each channel comprising the activity of several
neurons. We will demonstrate in this dataset that neuronal activity
significantly departs from a Gaussian distribution. Our data consist
of recordings from the visual cortices V1 and V4 of macaque monkeys
obtained by \citet{Chen22_77}. Fig.~\ref{fig:data-overview}a) shows
these neuronal recordings. A comprehensive overview about the data
and its preprocessing can be found in App.~\ref{sec:supplement-data}.
\citet{MoralesGregorio2024neural} study resting state recordings
from this dataset, where the monkey's eyes were tracked by a camera
(open or closed) while the monkey did not perform any task. They find
that these two states correspond to two distinct neural manifolds
using a Gaussian mixture model (GMM). We here examine these two manifolds
in more detail.

We showcase this dataset and its properties for an example recording
session in Fig.~\ref{fig:data-overview}. The data consist of 774
signal channels. For each channel, the multi-unit activity envelope
(MUAe) is $z$-scored over time. In Fig.~\ref{fig:data-overview}b),
the time evolution of the ten PCA variables (PCs) with the largest
eigenvalues are displayed alongside the label of the eye state. We
observe that certain PCs show a distinguishable behavior for the two
different eye states. Furthermore, the eigenvalues of the PCs decay
rapidly, indicating that a few variables explain most of the variance,
and thus supporting the hypothesis that the data lie on a low-dimensional
manifold. 

Inspecting Fig.~\ref{fig:data-overview}a) more closely, we note
that the distributions for some channels appear to have heavy tails
(having in mind that all channels are normalized to have unit variance).
This suggests that certain channels deviate clearly from Gaussianity,
a point we will quantify more systematically below using 4\textsuperscript{th}~order
cumulants.

The marginalized distributions in Fig.~\ref{fig:data-overview}c)
indicate that the data are neither Gaussian nor a mixture of two Gaussians.
While for some PCs, the marginalized distributions are close to a
Gaussian shape (e.g.\ PCs 5,7 \& 10), others deviate clearly from
a Gaussian (e.g.\ the first four PCs). The 2D marginalization of
the probability density along the first two PCs in Fig.~\ref{fig:data-overview}d)
appears to be bimodal, which is consistent with the ability of \citet{MoralesGregorio2024neural}
to fit the data with a GMM of two components. However, the overall
shape of the distribution differs markedly from what would be expected
for a mixture of Gaussians: instead of near-elliptical contours, the
probability density is concentrated within a well-defined triangular
region, suggesting strong deviations from Gaussian.

While these visualizations already suggest non-Gaussian statistics,
a more quantitative way to describe such deviations is through higher-order
cumulants, which also capture statistical dependencies in high-dimensional
spaces. For any Gaussian distribution, cumulants beyond second order
vanish; non-zero values of such cumulants can thus be used as an indicator
for deviations from Gaussianity. For this dataset, we observe significant
3\textsuperscript{rd} and 4\textsuperscript{th}~order cumulants
(Fig.~\ref{fig:data-overview}e-f). While higher-order cumulants
vanish for the close-to-Gaussian PCs 5, 7, and 10, they yield clearly
negative (blue) or positive (red) values for other PCs, in particular
PCs 1 and 2. The strongly positive values on the diagonal of the 4\textsuperscript{th}~order
cumulant may indicate heavy tails in these directions.

These findings provide concrete evidence of structured, non-Gaussian
statistics in the data. As we pointed out in the introduction, such
higher-order correlations are closely tied to the curvature of the
underlying manifolds. This connection highlights the need for a method
capable of capturing these structures in order to gain a more complete
understanding of the neural representation.

\section{Unsupervised learning of data distribution with Normalizing flows\protect\label{sec:normalizing-flows}}

In order to learn the underlying distribution of the neuronal activity,
we now employ Normalizing Flows (NFs) \citep{DInh15_1410,Dinh2016density,Kingma2018glow}
-- a class of invertible neural networks designed to learn complex,
high-dimensional probability distributions that is used in different
fields of machine learning. In the following paragraphs, we will introduce
NFs and our adaptations which enable them to simultaneously learn
a low-dimensional latent representation, characterize the geometry
of the data manifold, and provide interpretable access to its curvature
and higher-order correlations.

These neural networks define a bijective map 
\begin{eqnarray}
z & = & f_{\theta}(x)\label{eq:nf-definition}
\end{eqnarray}
from input data $x\in\mathbb{R}^{N}$ to latent variables $z\in\mathbb{R}^{N}$.
The inverse 
\begin{eqnarray}
x & = & f_{\theta}^{-1}(z)\label{eq:nf-inverse-definition}
\end{eqnarray}
maps back from the latent space to the data space. The network parameters
$\theta$ are learned so that the data distribution $p_{x}(x)$ is
mapped to a simple distribution $p_{z}(z)$ in the latent space. An
estimator $\hat{p}_{x;\theta}(x)$ for the true underlying distribution
$p_{x}(x)$ of the data can then be obtained as
\begin{eqnarray}
\hat{p}_{x;\theta}(x) & = & p_{z}(f_{\theta}(x))\,\left|\det\frac{\partial f_{\theta}(x)}{\partial x^{T}}\right|\,.\label{eq:likelihood}
\end{eqnarray}
We here focus on volume-preserving NFs \citep{Dinh2016density,Draxler2024universality},
where the building blocks of the networks are chosen such that the
determinant of the Jacobian is constant. Further details on the network
architecture can be found in App.~ \ref{subsec:Normalizing-flow-architecture}.

To find the estimator $\hat{p}_{x;\theta}$ which best represents
the data, we train the network by minimizing its negative log-likelihood
$\mathcal{L}_{\text{ll}}$ via 
\begin{eqnarray}
\mathcal{L}_{\text{ll}}(\theta) & = & -\frac{1}{N}\frac{1}{|X|}\sum_{x\in X}\log(\hat{p}_{x;\theta}(x))\label{eq:log_likelihood}
\end{eqnarray}
given the dataset $X$ \citep{DInh15_1410}. We normalize this loss
function by the number of dimensions of the dataset, i.e., the number
of channels $N$. In our application, this training reliably captures
the data distribution of the electrophysiological data, setting the
foundation for the extraction of meaningful and interpretable latent
variables in the next step. Details of the training procedure, including
optimizer settings, are described in App.~\ref{sec:sup-loss-function-training}.

\section{Extracting meaningful latent variables with reconstruction error
loss\protect\label{sec:reconstruction-error}}

\cprotect\paragraph{
\begin{figure}
\protect\begin{centering}
\protect\includegraphics{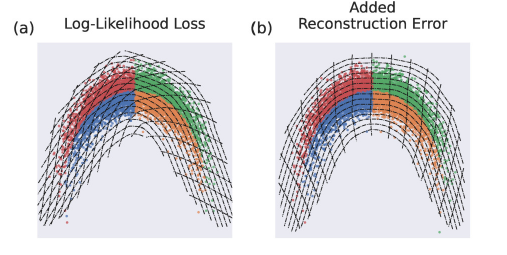}\protect
\par\end{centering}
\protect\caption{\protect\label{fig:GoC-data-latent}\textbf{Normalizing Flow with
reconstruction error learns inherent latent variables in a synthetic
task. }We illustrate the effect of of adding the reconstruction error
term using a synthetic dataset. Its topology naturally favors a description
in terms of two latent variables corresponding to the two main axes,
indicated here by different colors for the four quadrants of the data
cloud separated by the two main axes. We train an invertible network
$x=f_{\theta}^{-1}(z)$ on data points $x\in X$. In \textbf{(a)},
the network is trained by maximizing the log-likelihood of $x$ under
the model (Eq.~\ref{eq:log_likelihood}), using Gaussian distributed
latent variables $z\propto\mathcal{N}(0,\mathbb{I})$. In \textbf{(b)},
an additional reconstruction error term is included in the loss function
that encourages a small number of informative latent variables. For
both cases, we show grid lines corresponding to the axes system $f_{\theta}^{-1}(s\,e_{1}+t\,e_{2})$
when varying only one of the two latent variables $s,t\in\mathbb{R}$
at a time. Steps between neighboring lines are half a standard deviation
apart in the latent space. The lines are cut such that $\sqrt{t^{2}+s^{2}}\le4$.
By including the reconstruction error, the network correctly identifies
the two inherent latent variables of the dataset: In \textbf{(b)},
the grid lines align with the two main axes.}
\end{figure}
}

While log-likelihood maximization enables the NF to accurately model
the data distribution, it does not ensure that the latent variables
are interpretable or organized in a meaningful way. To address this,
we use an additional reconstruction error loss that enforces a hierarchical
structure among the latent variables, ordering them by their relevance
to data reconstruction. This hierarchy enables us to identify a small
subset of latent variables that effectively describe the underlying
data manifold.

To implement this, we map a data point $x$ to the latent space $z=f_{\theta}(x)$,
set all but the first $l$ dimensions of $z$ to zero, and reconstruct
the data via the inverse map:
\begin{eqnarray}
\tilde{x}_{\theta}^{:l}(x) & = & f_{\theta}^{-1}(\sum_{i=1}^{l}f_{\theta;i}(x)\hat{e}_{i})\,,\label{eq:reconstruct-single-point}
\end{eqnarray}
where $\hat{e}_{i}$ is the base vector of the $i$-th latent dimension.
If the data lies on a $l$-dimensional manifold, this reconstruction
will be exact, $\tilde{x}_{\theta}^{:l}(x)=x$, for all data points
$x\in X$. In general, the reconstruction error for $l$ dimensions
on the whole dataset is given by
\begin{eqnarray}
R_{l}(\theta) & = & \frac{1}{|X|}\sum_{x\in X}|x-\tilde{x}_{\theta}^{:l}(x)|^{2}\,.\label{eq:reconstruct-single-dim}
\end{eqnarray}
As the dimensionality is unknown, we average this reconstruction error
over the number of dimensions $l$ up to a cut-off $N_{r}$ to obtain
the full reconstruction error
\begin{eqnarray}
\mathcal{L}_{\text{rec}}(\theta) & = & \gamma_{\text{rec}}\frac{1}{N_{r}}\sum_{l=1}^{N_{r}}R_{l}(\theta)\,,\label{eq:L_rec}
\end{eqnarray}
where the prefactor $\gamma_{\text{rec}}$ is a hyperparameter. A
similar reconstruction error has been proposed by \citet{Bekasov2020ordering};
in contrast to their work, we here use equal weights $p_{l}=1/N$
for the reconstruction errors $R_{l}(\theta)$ in all $l$ dimensions.

The effect of this additional loss term can be seen in Fig.~\ref{fig:GoC-data-latent}
for a simple task. The task is constructed with two inherent latent
variables: one corresponding to the radius and one corresponding to
the phase. We visualize the data distribution in terms of drawn data
samples and the learned distribution in terms of grid lines corresponding
to the learned latent variables. Training a NF only on the negative
log-likelihood (Eq.~\ref{eq:log_likelihood}) in panel (a) yields
a good estimate of the data distribution. The latent variables of
the learned distribution, however, do not match the inherent latent
variables of the task. By including the reconstruction error (Eq.~\ref{eq:L_rec})
in the loss when training a NF (Fig.~\ref{fig:GoC-data-latent}b)),
the network learns these inherent latent variables: radius and phase.
For this task, we may even map the data to a lower-dimensional manifold:
setting the latent corresponding to the radius to zero,we obtain a
one-dimensional curve that lies in the middle of the data points that
still explains the data well. This two-dimensional example illustrates
the principle to obtain a low-dimensional latent space.

Although simple, the two dimensional example in Fig.~\ref{fig:GoC-data-latent}
captures the essence of our method: to learn arbitrary distributions
from data, align latent dimensions to interpretable axes, and thereby
identify lower-dimensional manifolds by omitting the least informative
dimensions.

\section{Multiple behavioral states can be captured by multimodal latent space\protect\label{sec:multimodality}}

\begin{figure*}
\includegraphics{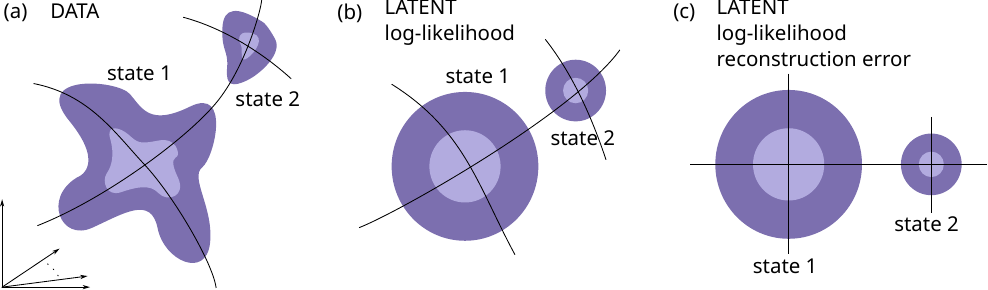}\caption{\protect\label{fig:net-overview}\textbf{Multimodal distributions
in data and latent space. (a)} A bimodal probability density in high-dimensional
data space. The two modes may correspond to two distinct system states,
such as different behavioral states.\emph{ }\textbf{(b)} Latent space
with bimodal distribution when trained only with the log-likelihood
loss term (Eq.~\ref{eq:log_likelihood}). \textbf{(c)} We observe
an alignment of sets of axes of the latent space to the behavioral
states due to the added reconstruction error (Eq.~\ref{eq:L_rec}).}
\end{figure*}

Most experimental datasets comprise data recorded within different
behavioral states. This is typically reflected in a data distribution
that has different modes or local maxima, as illustrated in Fig. \ref{fig:net-overview}a).
If the extent of each mode is not too large, its center can be regarded
as an archetypical data point that is representative of the corresponding
behavioral state. Since the Normalizing Flows we are using here are
volume-preserving, the number of modes is the same in the latent space
and in the data space \citep{Dinh2016density,Merger23_041033}.

To represent multimodal data distributions, we therefore use a Gaussian
mixture distribution in the latent space \citep{Izmailov2020semi},
illustrated in Fig.~\ref{fig:net-overview}b-c). For the mixture
components in latent space, we allow their centers to differ only
in the first $N_{l}$ dimensions -- the center coordinates of the
remaining $N-N_{l}$ dimensions are identical. Likewise, the variances
of the modes may differ only in the same first $N_{l}$ dimensions.
This strategy is needed to prevent overfitting, i.e., setting the
parameter $N_{l}$ too high, the network does not generalize well
to unseen data points. Both center coordinates and covariances are
learned, while $N_{l}$ is a hyperparameter that must be chosen a
priori. Maximizing the log-likelihood alone results in the situation
illustrated in Fig.~\ref{fig:net-overview}b): Neither of the two
latent variables alone distinguishes the switch from one state to
the other. Including the reconstruction error as an additional loss
term, in contrast, often led to an alignment of a set of latent dimensions
with the transition between behavioral states (Fig.~\ref{fig:net-overview}c)).
While this alignment is not guaranteed, it reflects a tendency for
the model to organize latent variables in an interpretable manner,
where changes between modes are associated with specific dimensions.
Details on how the means and the covariances of the latent components
are trained, together with their respective weights and on the decision
to have $N_{l}=10$ latent variables, in which the means and covariances
differ, can be found in App.~\ref{sec:sup-loss-function-training}
.

\section{Estimating the number of latent components of the data\protect\label{subsec:number-of-components}}

\begin{figure}[b]
\includegraphics{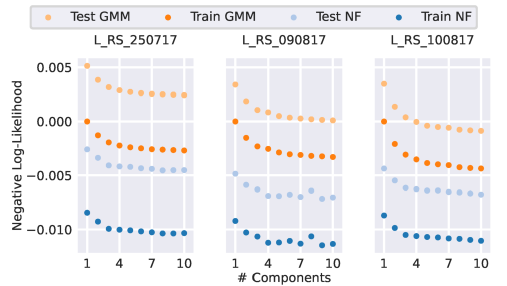}\caption{\protect\label{fig:training-comparison}\textbf{Accuracy of data representation
saturates with less than 10 latent components.} For three different
recording sessions, we show the negative log-likelihood per dimension
for both train and test data is shown as a measure of representation
quality. We compare Normalizing Flows (NFs) to linear Gaussian mixture
models (GMMs) to evaluate the advantage of learning a curved representation.
NFs show a saturation at 3 to 4 latent components. GMMs perform overall
worse for all numbers of components. An $80
$ train-test split is fixed within the analyses of a recording session.}
\end{figure}

\begin{figure*}
\includegraphics{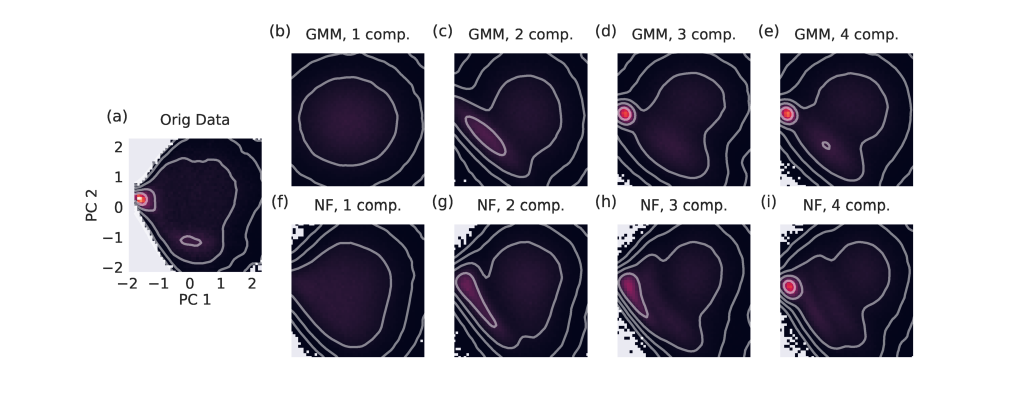}\caption{\protect\label{fig:component-comparison}\textbf{Effect of number
of latent components on the approximation capabilities: Marginalized
distributions} for the experimental data and learned models. \textbf{(a)
}Marginalized distribution of one session of experimental data (L\_RS\_250717)
along the first and second PCA directions. Light-gray isolines represent
equal likelihood. \textbf{(b--e)} Marginalized probability densities
for data sampled from Gaussian Mixture Models (GMMs) with increasing
numbers of latent components (1 to 4). \textbf{(f--i)} Corresponding
distributions for Normalizing Flows (NFs), also with 1 to 4 latent
components.}
\end{figure*}
 Having introduced our Normalizing Flow framework, we now apply it
to the neural activity data shown in Fig.~\ref{fig:data-overview}.
This requires choosing appropriate values for several hyperparameters,
the most critical of which is the number of Gaussian mixture components
in the latent space. This choice determines the model’s ability to
represent distinct behavioral modes within the data.

Each component can be interpreted as a flat manifold in latent space,
which is then mapped to a curved manifold in data space by the NF.
The number of components significantly determines the achievable accuracy
of the data description: for too few components, the description may
be sup-optimal and may miss relevant properties of the data; while
with too many components, the manifolds start to overlap and thus
become less interpretable.

To determine a suitable number of components, we train multiple instances
of the network with different numbers of mixture components. The log-likelihood
after training indicates the quality of the data description and is
shown in Fig. \ref{fig:training-comparison} for up to ten components.
For comparison, we also show results from fitting linear Gaussian
mixture models (GMMs). Their log-likelihood is overall worse than
that of the NFs, indicating a less accurate representation of the
data and thus a clear benefit of learning curved representations with
NFs over linear methods like PCA. While the log-likelihood of the
NFs monotonically increases with the number of components, it appears
to saturate after three or four. Thus, NFs yield not only a more accurate
representation of the data, but also require only a few mixture components.

To assess the learned data representations also qualitatively, we
show in Fig.~\ref{fig:component-comparison} the marginalized probability
density in the first two PCA variables of the data for different numbers
of mixture components in the latent space compared to the true marginalized
distribution. To express the strong contrast between the two small,
high-density regions and the vast, low-density region of the probability
density, multiple components are necessary. From this visualization,
we can also discern the reason for the smaller log-likelihood of GMMs
compared to the NFs: The linear GMMs fail to account for the triangular
cutoff of the probability density. Regions of high density, away from
the triangular corners, are similarly well represented by the linear
Gaussian mixture model and the representation continuously improves
when increasing the given number of components without overfitting.
The latter is expected, since a GMM with an arbitrary number of components
is able to represent any empirical distribution. In contrast, the
NFs are able to represent the triangular cutoff well even with few
latent components due to their non-linear mapping.

Based on these assessments, we consider Normalizing Flows with four
latent components for the remainder of this paper. Furthermore, we
focus on the first session (L\_RS\_250717) for the presentation of
the results, as the results from the other sessions are comparable
and consistent (see App.~\ref{sec:Supplementary-Figures}).

\section{Mixture components of neural activity correlate with behavioral states\protect\label{subsec:label-matching}}

\begin{figure}[b]
\includegraphics[width=1\columnwidth]{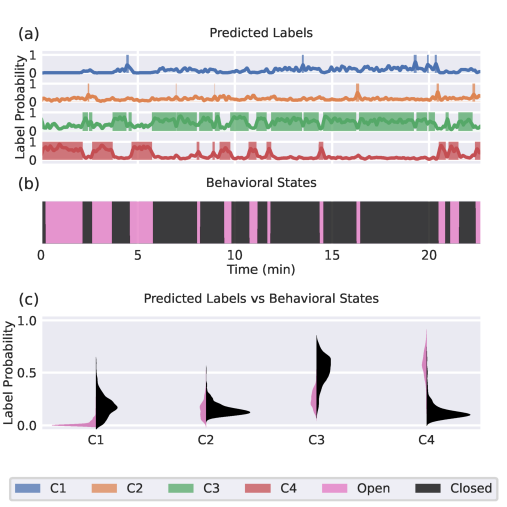}

\caption{\protect\label{fig:label-alignment}\textbf{Distinct mixture components
in the latent space represent behavioral states. (a)} For each of
the four components (C1--C4), smoothed (Gaussian filter, $\sigma=3\,\text{s}$),
normalized likelihoods are shown as solid curves. Labels are assigned
by selecting the component with the largest label probability at each
time point, indicated by background shading. \textbf{(b)} Ground-truth
behavioral states -- eyes open (pink) and eyes closed (black) --
are shown for direct comparison with the labels predicted in (a).
\textbf{(c)} The label probabilities are compared against the behavioral
states. For each of the components, smoothed histograms display the
distributions of label probabilities for the two behavioral states
(eyes open: left, eyes closed: right).}
\end{figure}

In the previous section, we found that four latent components yield
an accurate representation of the empirical data density for the electrophysiological
data. If the multimodality of the data is partly due to the animal
being in either of two different behavioral states -- eyes open or
eyes closed -- then we would expect some mixture components to correlate
with those states. And since the model has been trained without providing
behavioral labels, this would indicate that states are distinguishable
by structurally different neural activity.

To test this hypothesis, we compute the posterior probability of each
component $k$ given a neural data point $x$, denoted $p(k|x)$.
This is obtained via Bayes’ theorem:

\begin{eqnarray}
p(k|x) & = & \frac{p_{k}p(x|k)}{p(x)}=\frac{p_{k}p(x|k)}{\sum_{k'}p_{k'}p(x|k')}\label{eq:bayes-label}
\end{eqnarray}
Here, $p_{k}$\LyXZeroWidthSpace{} is the weight of the mixture component
$k$, and $p(x\mid k)$ is the likelihood of the data point under
that component. In a standard Gaussian mixture model, this likelihood
would simply be $\mathcal{N}(\mu_{k},\Sigma_{k};x)$. However, since
we use Normalizing Flows, we compute likelihoods using the change-of-variable
formula:

\begin{eqnarray}
\hat{p}_{x;\theta}(x|k) & = & \mathcal{N}(\mu_{k},\Sigma_{k};f_{\theta}(x))\,\left|\det\frac{\partial f_{\theta}(x)}{\partial x^{T}}\right|\,.\label{eq:likelihood-component}
\end{eqnarray}
where $f_{\theta}$\LyXZeroWidthSpace{} is the NF that maps data $x$
into the latent space. These component-wise likelihoods, together
with the priors $p_{k}$\LyXZeroWidthSpace , yield the normalized
posterior $p(k|x)$, which we interpret as the label probability for
each component at a given time point.

We denote the time-varying label probabilities as $p(k|t):=p(k|x(t))$,
and apply temporal smoothing with a Gaussian kernel ($\sigma=3\,\text{s}$)
to obtain smoothed probabilities $q(k|t)$. Fig.~\ref{fig:label-alignment}a)
shows these smoothed label probabilities as solid curves. To assign
a dominant component at each time point, we compute $\text{argmax}_{k}q(k|t)$,
which is indicated by background shading in the figure.

Comparing the dominant components in Fig.~\ref{fig:label-alignment}a)
with the true behavioral states in Fig.~\ref{fig:label-alignment}b),
we observe that components C4 and C3 correlate strongly with the eyes-open
and eyes-closed states, respectively. Components C1 and C2, which
have lower likelihoods across time, appear to account for residual
variability or transitions between dominant states.

Panel (c) further supports this relationship by directly comparing
the distribution of label probabilities across behavioral conditions.
Each component’s distribution is split by behavioral state, revealing
a clear preference: C4 shows high label probabilities predominantly
during eyes-open periods, while C3 peaks during eyes-closed periods.
Both these components are mostly indicative of the behavioral state.
In contrast, C1 and C2 exhibit in general low probabilities, with
C1 showing a clear preference for the eyes-closed state. These results
confirm that the latent components learned by the model reflect meaningful
behavioral distinctions in neural activity.

Since we observe this relation between mixture components and behavioral
states consistently in all recording sessions (see App.~\ref{sec:Supplementary-Figures}),
we conjecture that our proposed method generally allows extracting
latent components from neural data that correlate with behavioral
states encoded in the neural activity.

\section{Analytically tractable network approximation\protect\label{subsec:quadratic-approximation}}

The trained Normalizing Flow successfully captures complex neural
population activity, but the mapping it implements is hardly interpretable
due to the large number of trained parameters. To gain more insight,
we therefore need to simplify the mapping to make it more amenable
to interpretation. Whereas the analysis of previous sections relates
high-level behavior to mixture components, here the analysis relates
latent variables of a specific component to the geometric and statistical
properties of the neural activity manifold.

To achieve this, we approximate the learned mapping for each component
by a quadratic function. We found the quadratic to be a good balance
between a reduction in expressivity and the ease of interpreting latent
variables, all while retaining the latent space structure and the
ability to utilize the learned log-likelihood by the approximation's
link to its invertible origin.

For each latent component, we approximate the inverse mapping $f_{\theta}^{-1}(z)$
(Eq.~\ref{eq:nf-inverse-definition}) by a quadratic mapping from
the latent space $z$ to the data space $x$ as
\begin{eqnarray}
x & = & q(z)\label{eq:quad_net_mapping}\\
 & = & \sum_{\alpha}\biggl(\frac{1}{2}\sum_{i,j}A_{ij}^{\alpha}z_{i}z_{j}+\sum_{i}B_{i}^{\alpha}z_{i}+c^{\alpha}\biggr)\hat{e}^{\alpha}\,.\nonumber 
\end{eqnarray}
The upper indices $\alpha$ denote the dimensions of the data space
with $\hat{e}^{\alpha}$ being its base vectors. Instead, the lower
indices $i,j$ refer to dimensions in the latent space. The parameters
$c^{\alpha}$, $b_{i}^{\alpha}$\LyXZeroWidthSpace , and $A_{ij}^{\alpha}$\LyXZeroWidthSpace{}
represent the bias, the linear, and the quadratic terms, respectively.
We omit here an additional index for each component for the sake of
readability. Note that $A_{ij}^{\alpha}=A_{ji}^{\alpha}$\LyXZeroWidthSpace{}
due to symmetry.

This approximation serves as the simplest non-linear model, capturing
quadratic interactions among the latent variables $z_{i}$. Remarkably,
we find that the trained deep network can be well approximated with
this quadratic function. A qualitative intuition for the validity
of this approximation can be gained when comparing in Fig.~\ref{fig:cumulants}
the upper column (panels a-d) against the middle column (panels e-h),
where respectively the inverse map and its approximation are used.
For a quantitative evaluation of the fitting procedure see App.~\ref{sec:Sup-Quadratic-Approximation}.

The observation that the quadratic function serves as a good approximation
is non-trivial, because the space of functions that can be encoded
by a deep network is far larger. Whether this approximation holds
will in general depend on both the dataset and the network architecture;
for the neural recordings studied here it serves as a useful approximation,
yielding valuable insights into the neural code as shown in the subsequent
sections.

\section{Beyond Gaussian statistics of neural coordination\protect\label{subsec:cumulants}}

A better understanding of the neural code requires moving beyond 2\textsuperscript{nd}~order
(Gaussian) descriptions towards a full characterization of the statistical
structure of neural coordination. This is evident from the non-Gaussian
statistics found in the data (Fig.~\ref{fig:data-overview}). Indeed,
the presence of higher-order interactions can play a crucial role
in shaping population dynamics \citep{Montani2009impact,Ganmor2011sparse}.

\begin{figure*}
\includegraphics{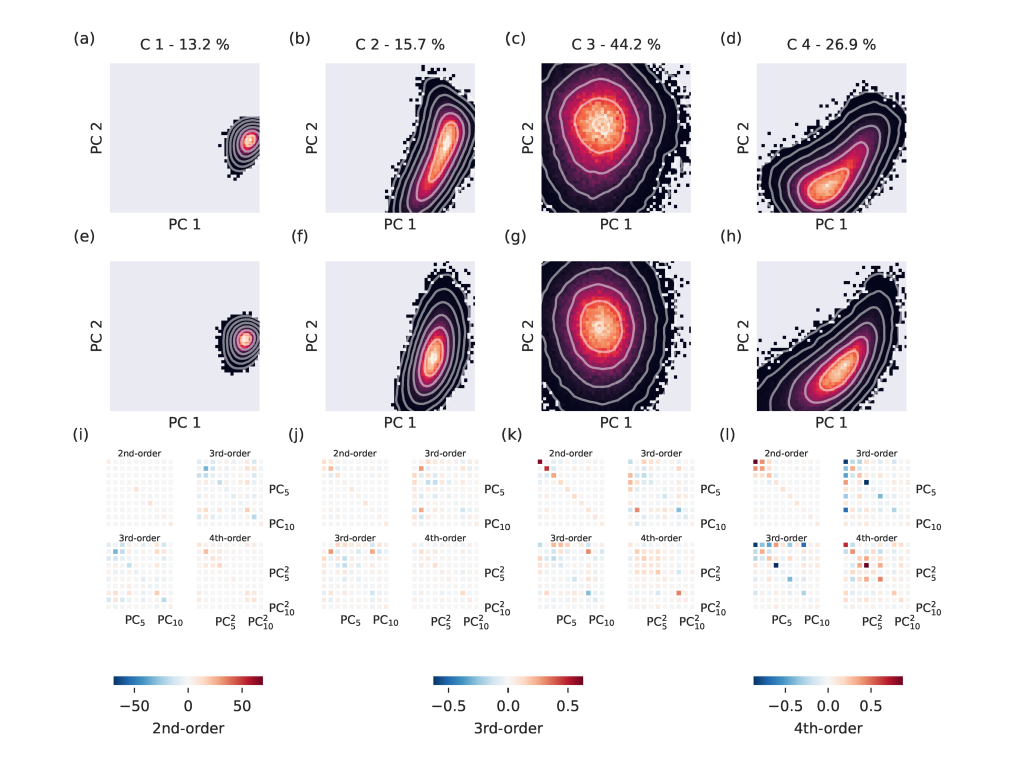}\caption{\protect\label{fig:cumulants}\textbf{Statistical characteristics
of the latent components: Approximate marginalized distributions}
for experimental data and learned models. \textbf{(a-d)} Marginalized
distributions of each latent component from the experimental session
(L\_RS\_250717), projected onto the first two PCA directions. Light-gray
contours indicate isolines of equal likelihood. \textbf{(e-h)} Marginalized
distribution of each component when the inverse mapping of the network
is approximated by a quadratic function (Eq.~\ref{eq:quad_net_mapping}).
\textbf{(i-l)} 2\protect\textsuperscript{nd} to 4\protect\textsuperscript{th}~order
cumulants computed from this approximation.}
\end{figure*}

As an illustration, in Fig.~\ref{fig:cumulants}a-d) we show the
probability densities that the network learned for each of the four
different components, marginalized onto the first two PCA variables.
Note that component 2 aligns to the eyes-open state, while the other
components mostly align to the eyes-closed state. Notably, component
C1 represents the high-likelihood mode in the tip of the triangular
structure, as observed in Fig.~\ref{fig:data-overview}. Component
C4 here stands out by showing a clear non-Gaussian structure.

The analytically tractable network approximation from the previous
section enables us to compute not just means and covariances, but
also 3\textsuperscript{rd} and 4\textsuperscript{th}~order cumulants
for each latent component. These quantities are derived from the characteristic
function of the approximate network mapping, providing a powerful
lens into the statistical structure of the data encoded by the model.
The statistics allows us to quantify the extent by which neural interactions
deviate from Gaussianity within each component.

We obtain the moment-generating function \citep{Gardiner09,Helias20_970}
from the quadratic approximation (Eq.~\ref{eq:quad_net_mapping})
of the network mapping for each component as
\begin{eqnarray}
Z(\bm{j}) & = & \frac{e^{\bm{j}^{T}\bm{c}}e^{\frac{1}{2}\bm{j^{T}}B(\mathbb{I}-\sum_{\alpha}A^{\alpha}j^{\alpha})^{-1}B^{T}\bm{j}}}{\det(\mathbb{I}-\sum_{\alpha}A^{\alpha}j^{\alpha})^{1/2}}\,.\label{eq:characteristic-function}
\end{eqnarray}
Alternatively we may consider the cumulant-generating function, given
by the logarithm of the moment-generating function. Derivations for
both the moment- and cumulant-generating functions are given in App.~\ref{sec:Sup-Cumulants}.

The resulting cumulants are an effective measure for non-Gaussianity
of distributions, since for Gaussian distributions, all cumulants
beyond 2\textsuperscript{nd}~order vanish. We show the cumulants
for all components in Fig.~\ref{fig:cumulants}e--h); we follow
here the same scheme as in Fig.~\ref{fig:graphical-abstract} by
showing all values of the 2\textsuperscript{nd}~order cumulant but
only a reduced set of values for the 3\textsuperscript{rd} and 4\textsuperscript{th}~order
cumulant.

As alluded to above, component C4 stands out by its pronounced 3\textsuperscript{rd}
and 4\textsuperscript{th}~order cumulants, marking it as clearly
non-Gaussian. Components C1 to C3 also show moderate higher-order
structure.

These results demonstrate that the components inferred by our model
exhibit distinct statistical signatures, with some showing strong
non-Gaussian features over several dimensions, highlighting the importance
of capturing higher-order structure to fully describe the neural population
activity across behavioral states. 

\section{Geometrical properties of the curved neural manifold \protect\label{subsec:curvature}}

To uncover the principles of neural coding, it is not only important
to understand the statistical structure of neural activity, but also
to characterize the geometric shape of the population activity manifold.
Neural activity patterns are often constrained to complex, non-linear
subspaces of the high-dimensional embedding space, and curvature is
a key indicator of such non-linearity. Having described the different
manifolds in our data from a statistical perspective, we now change
our perspective to a geometrical one. The geometrical perspective
offers complementary insight into the structure of how the brain might
organize neural states. It is also particularly relevant in light
of the higher-order statistical dependencies identified in the previous
section, as curvature can reflect underlying non-linear interactions
that are missed by 2\textsuperscript{nd}~order statistics.

Our method establishes a direct link between each learned latent component
and a corresponding sub-manifold in the high-dimensional neural activity
space. For the analyses presented here, we consider manifolds with
$N_{l}=10$ latent variables. We characterize the geometry of each
mixture component separately, by describing its curvature around its
point of maximum likelihood.

To do this, we build on the quadratic approximation (Eq.~\ref{eq:quad_net_mapping})
of the network mapping after training, from which we construct local
coordinate charts (see App.~\ref{sec:Sup-Curvature}). These charts
form a smoothly varying atlas that allows us to compute intrinsic
geometric quantities --- the scalar curvature, which measures the
overall average curvature, and the sectional curvature, which measures
how the manifold bends in a particular point relative to a two-dimensional
tangent plane. These concepts are central in differential geometry
and follow standard definitions found in e.g. \citet{Lee2012introduction,Lee2018introduction,Bar2010elementary,Robbin2022introduction}.
These two measures reveal how the neural representations deviate from
simple, flat geometries and provide insight into the complexity of
the underlying neural code.

\begin{figure}[b]
\includegraphics[width=1\columnwidth]{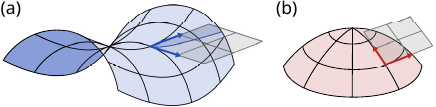}\caption{\protect\label{fig:sketch-curvature}\textbf{Geometrical interpretation
of data manifold curvature. }Tangent vectors (arrows) define the local
tangent space (gray) at a specific point on the data manifold. \textbf{(a)}
At a saddle point, the tangent space intersects the manifold, resulting
in negative sectional curvature. \textbf{(b)} For a spherical manifold,
the tangent space lies entirely outside the surface, indicating positive
sectional curvature.}
\end{figure}

For the sectional curvature, we consider pairs of latent variables
$i$ and $j$ and determine the curvature of the two-dimensional surface
they span on the data manifold. To do so, we first define a local
coordinate system on the data manifold in terms of tangent vectors
$e_{i}^{\alpha}$\LyXZeroWidthSpace . These tangent vectors describe
how perturbations in the latent variable $z_{i}$\LyXZeroWidthSpace{}
affect the data space $x^{\alpha}$ and are obtained from the quadratic
approximation of the network model as
\begin{eqnarray}
e_{i}^{\alpha} & = & B_{i}^{\alpha}+\sum_{j}A_{ij}^{\alpha}z_{j}\,.\label{eq:tangent_vectors}
\end{eqnarray}
Here $B_{i}^{\alpha}$\LyXZeroWidthSpace{} and $A_{ij}^{\alpha}$\LyXZeroWidthSpace{}
are the linear and quadratic coefficients, linking latent variables
to data dimensions. These vectors span the tangent space at a given
point on the data manifold (see \citep{Lee2012introduction}~Ch.3),
as illustrated in Fig.~\ref{fig:sketch-curvature}.

From these tangent vectors, we construct the metric tensor $g_{ij}$,
which defines the inner product on the tangent space and encodes local
geometric information such as distances and angles between directions
(see \citep{Lee2018introduction}~Ch.2):
\begin{eqnarray}
g_{ij} & = & \sum_{\alpha}e_{i}^{\alpha}e_{j}^{\alpha}\,.\label{eq:metric_tensor}
\end{eqnarray}
This metric tensor is essential in all further steps of the geometric
analysis and enters also in terms of its inverse $g^{ij}$.

Next, we introduce the projection operator $P^{\alpha\beta}$ onto
the normal space -- the space orthogonal to the tangent plane: 
\begin{eqnarray}
P^{\alpha\beta} & = & \delta^{\alpha\beta}-\sum_{n,p}e_{n}^{\alpha}g^{np}e_{p}^{\beta}\,,\label{eq:projector}
\end{eqnarray}
where $\delta^{\alpha\beta}$ is the identity matrix in data space.
This operator ensures that the curvature is calculated with respect
to deformations only perpendicular to the manifold.

Based on these definitions, we compute the sectional curvature $K(e_{i},e_{j})$
(see \citep{Lee2018introduction}~Ch.7), which quantifies how the
manifold bends in the plane spanned by the tangent vectors $e_{i}$\LyXZeroWidthSpace{}
and $e_{j}$ as

\begin{eqnarray}
K(e_{i},e_{j}) & = & \frac{\sum_{\alpha,\beta}P^{\alpha\beta}(A_{ii}^{\alpha}A_{jj}^{\beta}-A_{ij}^{\alpha}A_{ij}^{\beta})}{g_{ii}g_{jj}-(g_{ij})^{2}}\,.\label{eq:sectional_curvature}
\end{eqnarray}
We compute the scalar curvature $R$ by averaging over all sectional
curvatures at a point (see \citep{Lee2018introduction}~Ch.7). For
our model, this yields the expression:

\begin{eqnarray}
R & = & \sum_{j,l,c,m}\sum_{\alpha,\beta}g^{jl}g^{cm}P^{\alpha\beta}(A_{jl}^{\alpha}A_{mc}^{\beta}-A_{cj}^{\alpha}A_{ml}^{\beta})\,.\label{eq:scalar_curvature}
\end{eqnarray}
The scalar curvature $R$ provides a scalar invariant that summarizes
the intrinsic curvature of the manifold at a point. In two dimensions,
$R$ coincides with the Gaussian curvature and reflects whether the
sectional curvatures in the two PCA variable directions have equal
or opposite signs. Positive curvature corresponds to elliptic (spherical-like)
geometry, and negative curvature to hyperbolic (saddle-point-like)
geometry (see Fig.~\ref{fig:sketch-curvature}). In higher dimensions,
$R$ generalizes this concept by averaging the sectional curvatures
over all two-dimensional subspaces of the tangent space. A negative
value of $R$ indicates that, on average, the manifold bends away
in multiple directions, although not all sectional curvatures need
to be negative.

The detailed derivation of the expression for the scalar curvature
can be found in App.~\ref{sec:Sup-Curvature}.

\begin{table}
\begin{tabular}{|c|c|}
\hline 
 & Sectional curvature\tabularnewline
\hline 
\hline 
Component 1 & $-0.052$\tabularnewline
\hline 
Component 2 & $-0.011$\tabularnewline
\hline 
Component 3 & $-0.012$\tabularnewline
\hline 
Component 4 & $-0.097$\tabularnewline
\hline 
\end{tabular}\caption{\textbf{\protect\label{tab:Scalar-curvature}Scalar curvature} for
all four components.}
\end{table}

\begin{figure*}
\includegraphics{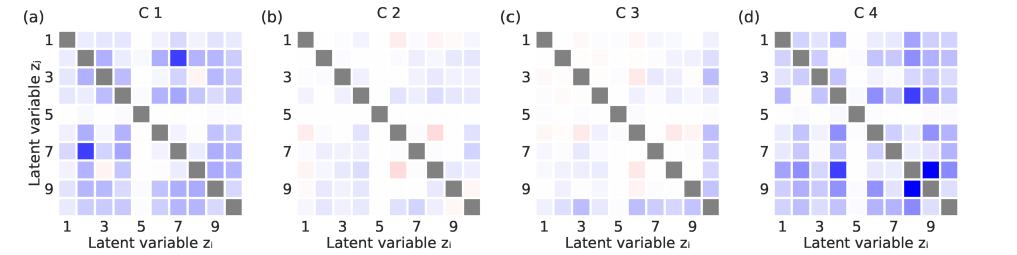}\caption{\protect\label{fig:sectional-curvature}\textbf{Geometrical characteristics
of the latent components. }Panels\textbf{ (a-d)} show the sectional
curvature for each of the four components of the latent distribution.
These values are computed from the quadratically approximated inverse
network. Red indicates positive curvature, while blue indicates negative
curvature. The color scale is consistent across all panels, with a
range defined by the values reported in Table~\ref{tab:Scalar-curvature}
for the scalar curvature (i.e.\ the mean of sectional curvatures).}
\end{figure*}

We measure the scalar curvature in Table~\ref{tab:Scalar-curvature}
for the different latent components: Component C4 exhibits the most
pronounced negative curvature (\textminus 0.097), followed by component
C1 (\textminus 0.052), both indicative of saddle-like manifold structures.
In contrast, components C2 and C3 show a nearly flat geometry with
a weak negative curvature (\textminus 0.011 / \textminus 0.012), suggesting
only minor deviation from flatness. As we saw in Fig.~\ref{fig:label-alignment},
component C3 is mostly aligned to the eyes-closed state. It is also
a component with moderate higher-order correlations (Fig.~\ref{fig:cumulants}).
Here we now observe that it is also a component lying on a near-flat
manifold. In contrast, component C4, aligning mostly to the eyes-open
state (Fig.~\ref{fig:label-alignment}) corresponds to the most curved
sub-manifold. This component also shows the strongest deviations from
Gaussianity (Fig.~\ref{fig:cumulants}). The two components which
do not align clearly to one of the states but rather account for transitions
between or periods in-between one of the states (Fig.~\ref{fig:label-alignment})
show a saddle-point-like (C1) and a near-flat (C2) geometry. 

A more detailed picture emerges from the sectional curvature matrices
shown in Fig.~\ref{fig:sectional-curvature}. Here, the curvature
contributions from different pairs of latent variables are visualized.
For components C1 and C4, the sectional curvature is predominantly
negative (blue-shaded entries), supporting the interpretation of a
globally saddle-point-like structure. In components C2 and 3, however,
the curvatures are more balanced, with both weakly positive and negative
values, aligning with its near-zero scalar curvature and suggesting
a less distinctly curved manifold.

These results suggest that curvature is not confined to a few exceptional
latent variables but is distributed across the entire latent space.
In other words, the manifold's curvedness reflects joint interactions
among multiple latent variables rather than isolated effects of a
few variables.

In particular, we observe that the component exhibiting the strongest
higher-order statistical correlations lies on the most curved manifold.
This alignment between statistical complexity and geometric structure
supports our claim that understanding neural coordination requires
tools capable of capturing both non-Gaussian statistics and non-linear
geometry; where our method extracts both aspects from a joint framework.

\section{Discussion\protect\label{sec:Discussion}}

This study presents an integrated framework designed to analyze complex
neural population activity, focusing on three key objectives: (1)
to extract coordinated activity across neurons, described both statistically
via correlations and geometrically as manifolds; (2) to identify a
compact set of latent variables that capture this structure and potentially
relate to behavior; and (3) to provide interpretable and analytical
characterizations of both the statistical properties -- using the
characteristic function -- and the geometric features -- providing
a set of charts and measuring the curvature of the manifold.

To achieve these objectives, we propose a novel method utilizing
Normalizing Flows (NFs) that describes how coordinated neural activity
represents information. The method provides insights from both a statistical
and a geometrical perspective, both of which rely on first training
an invertible neural network in an unsupervised manner on neural data.
An analytical description of the learned network mapping then enables
us to compute both the statistical cumulants -- of arbitrary order
-- and the geometric curvature of the original data.

Our approach merges the expressive capabilities of neural networks
with the clarity provided by straightforward analytic expressions.
As a result, it enables the extraction of non-Gaussian statistical
structure, non-linear dimensionality reduction, and the identification
of latent variables aligned with behavior. Applied to visual cortical
areas V1 and V4 in macaques, our method reveals interpretable latent
components that align with behavioral states, each associated with
distinct non-Gaussian statistics and curved sub-manifolds in neural
state space.

The invertibility of Normalizing Flows implies that they provide
explicit expressions for the likelihood of the data under the model:
they can therefore be trained using the true likelihood. This contrasts
with other architectures such as variational autoencoders \citep{Kingma2013auto},
where the likelihood is only implicit and therefore cannot be analyzed
directly. The use of an additional loss function allows us to distinguish
between latent variables that span the neural manifold and others
that are classified as noise dimensions. In addition, by choosing
a multimodal latent distribution, we may distinguish between the different
behavioral states in the neural code. Although the training is unsupervised,
the resulting representations still align with the original behavioral
states observed in the experiment -- indicating that our method can
identify behavioral states solely from observed recordings.

We examine the statistical and geometric properties of the neural
recordings of V1 and V4 in macaque recorded by \citet{Chen22_77}.
Statistically, the presence of 3\textsuperscript{rd} and 4\textsuperscript{th}~order
cumulants reveals strong non-Gaussian structure in the data, reflecting
the non-linear nature of interaction in the neural code. Geometrically,
the curvature analysis uncovers predominantly saddle-point-like manifolds,
characterized by consistently negative scalar and sectional curvature.
This reflects a globally organized, rather than fragmented, structure
of the neural code. Our framework enables a rigorous comparison of
these sub-manifolds across latent dimensions, offering a detailed
understanding of how they contribute to the overall shape and coordination
of neural activity.

Our curvature-based analysis connects to a growing body of work exploring
manifold geometry in both theoretical and applied contexts. On the
theoretical side, recent developments in diffusion geometry provide
tools to quantify curvature, tangent structures, and dimensionality
directly from data \citep{Jones2024manifold}, complementing foundational
treatments in Riemannian geometry and its application to deep generative
models \citep{Shao2018riemannian}. Curvature has also emerged as
a powerful analytic tool in neuroscience. For example, \citet{Acosta2023quantifying}
estimate extrinsic curvature to characterize the shape of neural manifolds,
focusing primarily on synthetic data and including only a small-scale
application to experimental place cell recordings. Their method however
focuses on a pure geometric analysis, and does not try to describe
higher-order statistical structure such as cumulants. In contrast,
our approach learns a full probability density over population activity,
enabling the extraction of higher-order correlations and a richer
description of neural variability. Furthermore, we apply our method
to large-scale recordings involving approximately 800 simultaneously
recorded neurons, providing a population-level analysis of manifold
geometry. Additionally, curvature-based network methods have revealed
differences in structural \citep{Farooq2019network} and functional
connectivity \citep{Weber2017curvature}. Further, curvature plays
a critical role in statistical analysis on manifolds, such as determining
the uniqueness of geometric medians \citep{Fletcher2009geometric}
or improving dimension estimation \citep{Gilbert2023pca}. While these
diverse efforts underscore the value of curvature as a geometric descriptor,
our approach distinctively integrates curvature-based geometric insights
with a statistical model of correlations while providing a simple,
interpretable structure.

A number of earlier theoretical contributions have emphasized the
importance of correlated variability in shaping population codes.
The review by \citet{Azeredo2021geometry} framed this in terms of
the geometry of information coding. Related theoretical work has further
highlighted how specific forms of correlation can impact information
encoding \citep{Morenobote14_1410,Kohn2016correlations}, and how
population activity may be better understood within a probabilistic,
inference-based framework \citep{Pouget2003inference}. Our framework
provides a concrete implementation of these ideas by learning both
the statistical correlations and the geometry of neural manifolds
from data.

Building on this foundation, recent studies have highlighted the inherently
non-linear nature of these manifolds, especially during behaviorally
rich tasks. For instance, multiple studies \citep{Gallego17_978,Gallego18_1,Gallego20_260,Safaie2023preserved,Altan2023low,Fortunato2024nonlinear}
have shown across multiple species and brain regions that low-dimensional
neural manifolds are not merely a useful abstraction, but that their
non-linear curvature and structure carry functional relevance. (For
a review refer to \citep{Mitchell2023neural}.) These findings support
the relevance of our approach, where we also go beyond linear assumptions
and explicitly extract the curved geometry of neural activity. A related
perspective is offered by \citet{Chung2021neural}, who advocate for
population-level geometric analyses as a powerful tool to study both
biological and artificial networks. Most directly, our work complements
recent findings by \citet{Chou2025geometry}, who demonstrated that
the curvature of neural manifolds is tightly linked to the efficiency
of representational untangling during task performance. Our work contributes
to this growing effort by providing an interpretable and flexible
framework for identifying such curved manifolds and linking their
geometry to meaningful aspects of neural computation.

To enable both the statistical and geometric analysis presented above,
we approximate the inverse of the learned network mapping by a quadratic
function. While this choice may not seem obvious, we show empirically
in App.~\ref{sec:Sup-Quadratic-Approximation} that the statistics
of the approximation closely match those of the network, even as we
go to higher-order, off-diagonal components. This quadratic form allows
us to compute statistical descriptors such as cumulants analytically,
as well as to characterize geometric features such as curvature from
the same functional representation. The resulting unified description
enables rigorous comparisons across latent dimensions and behavioral
states.

The increasing availability of large-scale neural recordings has
created a pressing need for methods that can link brain-wide activity
to computation and behavior \citep{urai2022large}. Addressing this
challenge requires approaches that are both theoretically grounded
and computationally scalable. With the targeted extensions to Normalizing
Flows introduced in this work, our method is designed to efficiently
model high-dimensional neural population activity. We applied it to
recordings with approximately 800 simultaneously recorded channels,
and its architecture is readily scalable to even larger datasets,
such as those obtained from Neuropixels probes with thousands of channels.
This makes our framework a timely and adaptable tool for extracting
interpretable structure from large-scale neural data.

Scaling to such high-dimensional recordings naturally raises concerns
about computational cost and model complexity. To address this, we
adopt an architectural strategy similar to that of \citet{Cramer2022principal},
where only a subset of the input dimensions are processed by the flow.
Specifically, after a fixed linear transformation based on the input
covariance, we pass only the top $N_{n}$ PCA variables through the
flow, while the remaining dimensions are directly passed to the latent
space. This design keeps the network size manageable and ensures efficient
training and evaluation. However, it comes with the trade-off that
potentially informative structure in the discarded dimensions is not
learned by the NF, which may limit the achievable log-likelihood.

An important constraint for applying Normalizing Flows (NFs) to other
datasets is the inability of NFs to model perfectly noise-free dimensions,
where the log-likelihood would diverge \citep{Loaiza2022diagnosing}.
In practical applications such as ours, this is not a concern due
to the inevitable presence of noise in electrophysiological recordings.
However, this limitation should be carefully considered when applying
our method to synthetic or near noise-free data.

In our implementation, we further constrain the architecture by selecting
all flow blocks to be volume-preserving \citep{Dinh2016density}.
While this is not a fundamental requirement of NFs, it is an explicit
design choice. Volume preservation prevents the network from learning
diverging mappings and contributes to more stable training dynamics,
as noted by \citep{Behrmann2021understanding}. Although this restriction
may limit model expressivity \citep{Draxler2024universality}, we
compensate this by using a more sophisticated latent space, which
also allows the separation of different behavioral states.

Some of the drawbacks described are inherent to NFs, and may not
be present in alternative architectures. Among deep generative models,
variational autoencoders (VAEs) \citep{Kingma2013auto} represent
a natural alternative to Normalizing Flows for modeling high-dimensional
neural activity in an unsupervised and probabilistic manner. While
VAEs do not provide exact likelihoods and require both an encoder
and a decoder network, they enable flexible non-linear dimensionality
reduction and generative modeling by optimizing a variational bound
(ELBO). Extensions such as $\beta$-VAE \citep{Higgins2017beta} and
similar VAE adaptations \citep{Hsu2017unsupervised,Kim2018disentangling,Chen2018isolating,Sarhan2019learning,Zhu2020learning,Zhu2021where}
promote disentangled and interpretable latent spaces through regularization
schemes, while more recent work has explored structured latent spaces
\citep{Connor2021variational}, mixtures of local VAEs for manifold
learning \citep{Alberti2024manifold}, and geometry-aware training
procedures \citep{Rey2019diffusion,Koehler2021variational}.

There appears to be no fundamental obstacle to extending our conceptual
approach to such architectures. The key methodological steps -- including
approximating the decoder with a quadratic function and using it to
compute characteristic functions, cumulants, and curvature tensors
-- could in principle also be applied to VAEs or similar architectures.
As long as the decoder admits a suitable analytic approximation, the
statistical and geometric descriptors introduced in this work remain
valid. This highlights that our core contributions are not specific
to NFs, but rather propose a general strategy for latent-variable
models.

Looking ahead, we identify several promising directions for extending
this work: applying the framework to larger and more diverse datasets,
incorporating time-varying manifold structures, analyzing multimodal
sensory representations, and refining the structure of the latent
space.

Although we here apply the framework to electrophysiological recordings
in the macaque visual cortex, its design is fully general. It can
be readily applied to neural population data from other brain regions
or species, offering a principled way to extract and interpret latent
structure in high-dimensional neural recordings. A natural next step
is to apply our approach to more diverse and complex electrophysiological
datasets. This will allow us to probe whether similar structured manifolds
emerge in neural systems e.g.\ involved in decision-making, motor
control, or memory. Likewise, datasets involving more naturalistic
or complex behaviors -- such as goal-directed navigation or learning
-- could reveal how neural manifolds change for more complicated
tasks. The framework is also applicable across species, enabling cross-species
comparisons of neural coding principles.

The modeling of time-varying manifolds, in which the statistical
and geometric structure of each latent component is allowed to evolve
over time, would be particularly relevant for very long recordings
or experiments conducted over multiple days, where animals may gradually
become more familiar with the task or experimental environment. In
such cases, tracking slow drifts or systematic changes in the manifold
structure or higher-order cumulants could reveal how internal models,
attention, or behavioral strategies are reshaped over extended timescales.

In multimodal perception, neural responses to combined stimuli --
such as visual and auditory inputs, or different features within the
same modality -- may superimpose in a non-linear manner, rather than
summing linearly. Our approach could help disentangle these complex
interactions by revealing how diverse sensory components are jointly
embedded and structured within the neural manifold.

Finally, further development of the latent space structure may enhance
the model’s flexibility. Our present model uses a Gaussian mixture,
which provides a clear and interpretable decomposition of the data
into distinct components. While being effective for many datasets,
certain processes might be better represented using circular latent
structures -- for instance, periodic variables seen in orientation
tuning or cyclic movements like gait -- wherein their underlying
phases exhibit inherent circularity. Alternatively, mixtures of more
flexible distributions -- rather than only Gaussians -- could yield
fewer components with the same expressive power. Incorporating such
structures into the model's latent space would allow it to represent
a broader class of neural dynamics while preserving interpretability.

To summarize, our approach provides a powerful, general tool for
revealing latent variables, statistical dependencies, and geometric
structure in high-dimensional neural data, setting the stage for future
advances in neural manifold analysis across brain areas, behaviors,
and species.

\section*{Code availability}

We will upload the code repository for the final version of this manuscript
as a zenodo-archive.

\section*{Acknowledgments}

This project is funded by the Deutsche Forschungsgemeinschaft (DFG,
German Research Foundation) - 368482240/GRK2416; and by the German
Federal Ministry for Education and Research (BMBF Grant 01IS19077A
to Jülich and BMBF Grant No. 01IS19077B to Aachen). SN acknowledges funding from Israel
Science Foundation grant 1442/21. CM acknowledges funding from Next
Generation EU, in the context of the National Recovery and Resilience
Plan, Investment PE1 -- Project FAIR “Future Artificial Intelligence
Research” (CUP G53C22000440006). The authors gratefully acknowledge
the computing time granted by the JARA Vergabegremium and provided
on the JARA Partition part of the supercomputer JURECA at Forschungszentrum
Jülich (computation grant JINB33). 

We thank the authors of \citet{Chen22_77} for publicly providing
the electrophysiological data from their macaque monkey experiments.
We also thank Aitor Morales-Gregorio and Anno C. Kurth for suggesting
this dataset, advice for the preprocessing and fruitful discussions.

\newpage{}

\section*{}

\appendix

\section{ Preprocessing of Data\protect\label{sec:supplement-data}}

We used publicly available 1024-channel electrophysiological recordings
from macaque V1 and V4 during resting state, recorded from \citet{Chen22_77}.
The data were acquired using 16 Utah arrays (1.5\,mm electrodes; Blackrock
Microsystems), with 14 arrays in V1 and 2 in V4, targeting primarily
layer 5. Preprocessing followed the procedure detailed in \citet{MoralesGregorio2024neural},
including the extraction of the multi-unit activity envelope (MUAe)
through high-pass filtering (500\,Hz), rectification, low-pass filtering
(200\,Hz), and downsampling to 1\,kHz, with notch filtering at 50,
100, and 150\,Hz to remove line noise. Electrodes with signal-to-noise
ratio below 2 or identified as contributing to high-frequency crosstalk
(following \citet{MoralesGregorio2024neural}) were excluded.

For the current analysis, we further downsampled the MUAe to 100\,Hz
using FIR-filter-based decimation, trimmed edge samples to remove
boundary effects, and z-scored each channel independently. The resulting
normalized signals across all arrays were concatenated into a single
matrix, forming the input for subsequent analysis.

\section{ Normalizing Flow Architecture\protect\label{subsec:Normalizing-flow-architecture}}

The Normalizing Flow learns the distribution of single time-points
(for all channels together) of the MUAe recordings. The architecture,
which is implemented in pytorch, begins with a single PCA-like transformation
inspired by \citet{Cramer2022principal}, which projects the input
onto the eigenbasis of its covariance matrix and normalizes each component
to unit variance. Only the $N_{n}=70$ largest principal components
are further processed by the flow, while the remaining dimensions
bypass it and are directly passed to the output, that is the latent
space. The flow comprises 10 volume-preserving blocks, each consisting
of a linear all-to-all mixing layer followed by an additive coupling
layer \citep{DInh15_1410}, using ELU activations and a hidden dimension
of 128. All components are invertible, allowing exact likelihood computation
and efficient sampling.

\section{ Training Procedure\protect\label{sec:sup-loss-function-training}}

The training process consists of pre-training and main training. Prior
to training, model parameters are initialized to promote numerical
stability and effective convergence. The single PCA-like transformation
projects the input onto the eigenbasis of its covariance matrix and
normalizes it to unit variance, based on the training data. This layer
stays constant over the training. The linear all-to-all and the additive
coupling layers are initialized using orthogonal matrices for the
linear layers and and biases are sampled uniformly from the interval
$[-1/N_{n},1/N_{n}]$, where $N_{n}$ is the number of dimensions
that is processed by the main building blocks. This initialization
ensures proper signal propagation and variance preservation through
the network.

For the initial latent distribution, a Gaussian Mixture Model (GMM)
with 4 components is fit to a projection of the training data after
the PCA-like transformation into a $N_{l}=10$-dimensional latent
space. We use the same value here as for $N_{r}=10$ (see Eq.~\ref{eq:L_rec}).
These low values are chosen so that the analysis of cumulants and
curvature can be displayed more clearly.

The GMM is implemented using scikit-learn's \emph{GaussianMixture}
model, and is refit at the beginning of each main-training round based
on the current latent output. Each component $k$ of the GMM is defined
by a weight $p_{k}$, a mean vector $\mu_{k}$ and a covariance matrix
$\Sigma_{k}$. These parameters are estimated by the GMM fitting procedure.
By design, $\mu_{k}$ is non-zero only in the first $N_{l}$ dimensions,
and $\Sigma_{k}$ is constrained to deviate from the identity only
in the top-left $N_{l}\times N_{l}$ block. These constraints also
hold for the refitting in the main training.\LyXZeroWidthSpace{}

In the pre-training stage, the network is trained for 100 epochs using
the standard likelihood loss function (Eq.~\ref{eq:log_likelihood}). 

The main-training consists of 20 rounds of 5 epochs each. Here, the
reconstruction error loss (Eq.~\ref{eq:L_rec}) is added. The weight
is chosen to be $\gamma_{rec}=0.02$. At the beginning of each round,
a new GMM is refit on the current latent output. The network parameters
are continuously trained throughout the training rounds.

Optimization in both phases uses the Adam optimizer with a learning
rate of 0.001, betas (0.9, 0.999), epsilon of 1e-8, and no weight
decay. A batch size of 10,000 time points is used throughout training.

\section{ Validity of Quadratic Approximation\protect\label{sec:Sup-Quadratic-Approximation}}

In this appendix, we describe the procedure used to approximate the
inverse mapping of each latent component by a quadratic function and
validate it. This approximation is intended to simplify the interpretation
of the complex mapping implemented by a trained Normalizing Flow.

The procedure starts from the latent components which are a result
of the training procedure (see App.~\ref{sec:sup-loss-function-training}).
Each component $k$ is characterized by its mean~$\mu_{k}$ and covariance~$\Sigma_{k}$.
Samples $z$ drawn from these components are mapped back to data space
via the inverse flow, $x=f_{\theta}^{-1}(z)$.

For each mixture, we perform a quadratic fit to approximate the inverse
mapping from latent space $z$ to data space $x$. The steps involved
are:

\paragraph{Feature Construction:}

For each latent mixture, features are constructed by including both
linear terms ($z_{i}$\LyXZeroWidthSpace ) and quadratic terms ($z_{i}z_{j}$\LyXZeroWidthSpace )
to be able to fit all coefficients of Eq.~\eqref{eq:quad_net_mapping}.
The feature construction is achieved using \emph{PolynomialFeatures}
from scikit-learn.

\paragraph*{Model Fitting:}

A \emph{LinearRegression} model is used to fit these polynomial features
against the corresponding data mixtures. This fitting captures both
linear and quadratic interactions among latent variables. The parameters
obtained from this fitting process---bias $c^{\alpha}$, linear terms
$B_{i}^{\alpha}$\LyXZeroWidthSpace , and quadratic terms $A_{ij}^{\alpha}$\LyXZeroWidthSpace ---are
crucial for understanding how changes in latent variables affect neural
population activity.

Empirical analysis (see Figs.~\ref{fig:scatter-1}, \ref{fig:scatter-2},
and \ref{fig:scatter-3}) reveals that the cumulants up to order 4
are well approximated by the quadratic function across all components:
although deviations appear, especially for high-order off-diagonal
components, the approximated statistics remain strongly correlated
with the true ones. This observation is significant as it indicates
that despite the reduced complexity of the quadratic model compared
to the full network, it retains essential statistical properties of
neural activity. The close alignment between original and approximated
cumulants suggests that key interactions among latent variables are
preserved, providing confidence in using this simplified model for
further interpretation and analysis.

\begin{widetext}

\begin{figure}[H]
\includegraphics{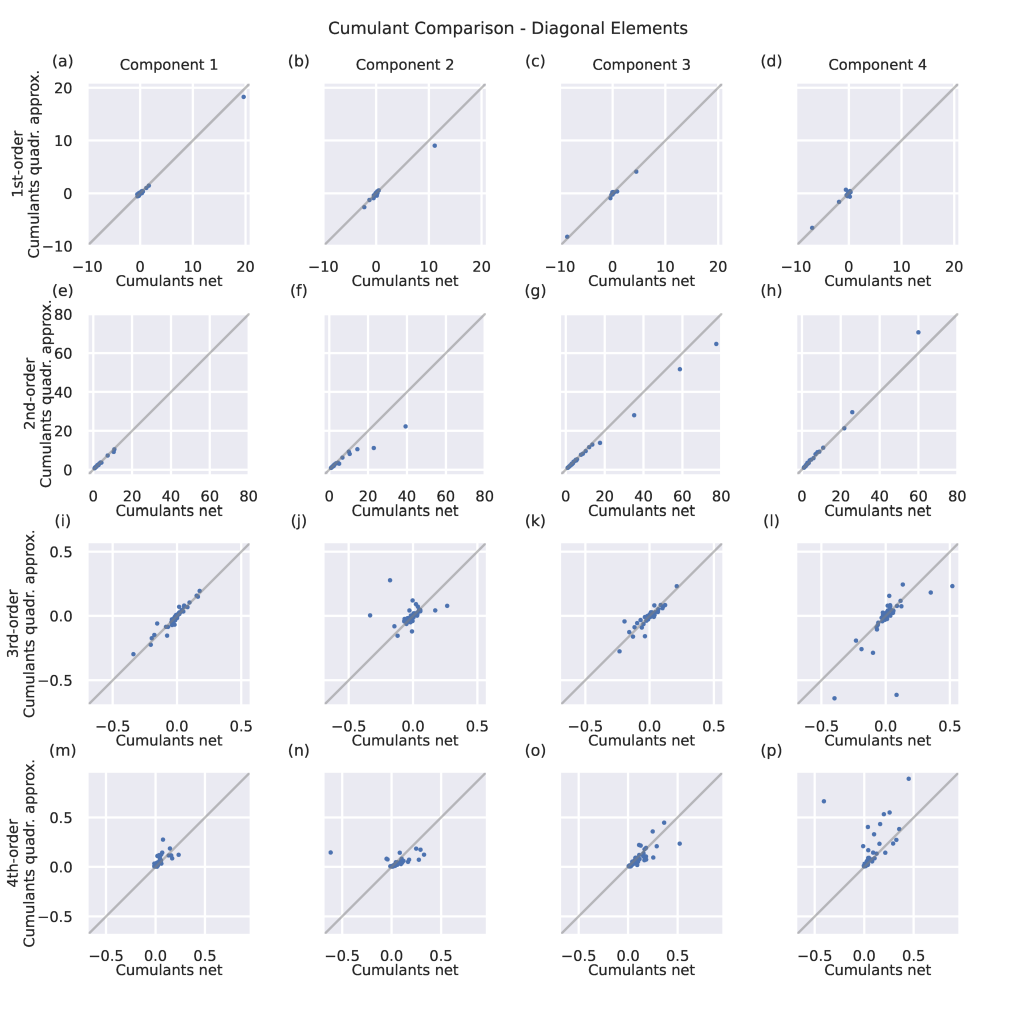}\caption{\protect\label{fig:scatter-1}Comparison of diagonal elements of cumulants
between the original inverse mapping of the network $f_{\theta}^{-1}$
(Eq.~\ref{eq:nf-inverse-definition}) and its quadratic approximation
(Eq.~\ref{eq:quad_net_mapping}) for each of the four components.
Each row corresponds to a different order of cumulants, from 1\protect\textsuperscript{st}~order
(panels \textbf{a-d)} to 4\protect\textsuperscript{th}~order (panels
\textbf{m-p)}. Each column corresponds to one of the four latent components.}
\end{figure}

\begin{figure}[H]
\includegraphics{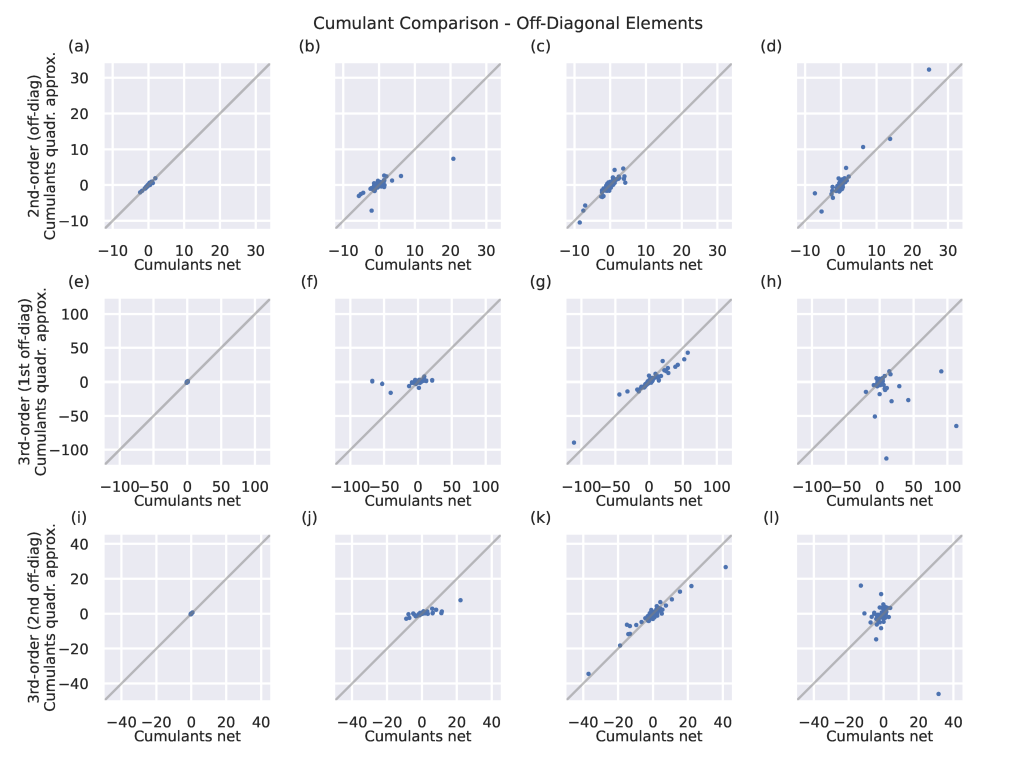}\caption{\protect\label{fig:scatter-2}Comparison of off-diagonal elements
of cumulants between the original inverse mapping of the network $f_{\theta}^{-1}$
(Eq.~\ref{eq:nf-inverse-definition}) and its quadratic approximation
(Eq.~\ref{eq:quad_net_mapping}) for each of the four components.
The first row shows off-diagonal elements for the 2\protect\textsuperscript{nd}~order
cumulants. The second and third row show 1\protect\textsuperscript{st}
and 2\protect\textsuperscript{nd}~order off-diagonal elements for
3\protect\textsuperscript{rd}~order cumulants. Each column corresponds
to one of the four latent components.}
\end{figure}

\begin{figure}[H]
\includegraphics{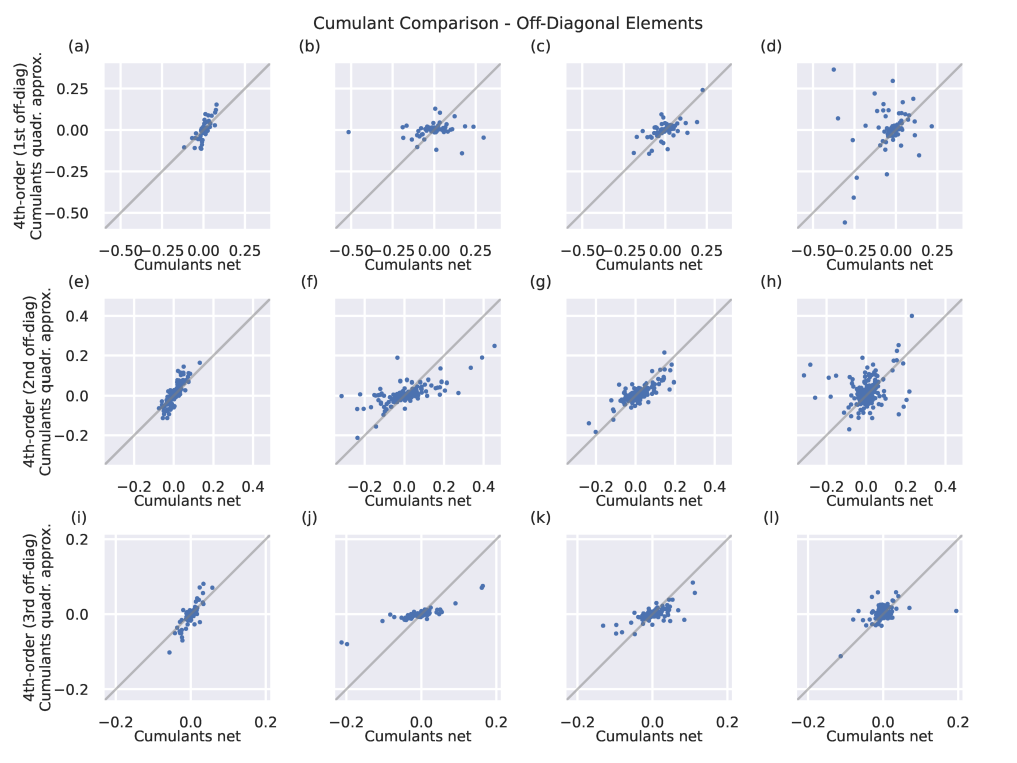}\caption{\protect\label{fig:scatter-3}Comparison of off-diagonal elements
of cumulants between the original inverse mapping of the network $f_{\theta}^{-1}$
(Eq.~\ref{eq:nf-inverse-definition}) and its quadratic approximation
(Eq.~\ref{eq:quad_net_mapping}) across four components for 4\protect\textsuperscript{th}~order
cumulants. Each row corresponds to a different degree of off-diagonality
of cumulant, from 1\protect\textsuperscript{st}~order off-diagonal
(panels \textbf{a-d)} to 3\protect\textsuperscript{rd}~order off-diagonal
(panels \textbf{i-l)}, demonstrating the effectiveness of the quadratic
fit.}
\end{figure}

\end{widetext}

\section{Characteristic function and Cumulants\protect\label{sec:Sup-Cumulants}}

To compute the cumulants of the data variables $x^{\alpha}$, we approximate
the trained Normalizing Flow $f_{\theta}^{-1}$ by a second-order
(quadratic) function of the latent variables $z$ \eqref{eq:quadratic_approx}.
For each mixture component, the latent variables follow a multivariate
Gaussian distribution. This structure allows for the analytical evaluation
of the moment generating function $Z(\boldsymbol{j})$ (characteristic
function), and from it, the cumulant generating function $W(\boldsymbol{j})=\ln\,Z(\boldsymbol{j})$.
By expanding $W(\boldsymbol{j})$ in powers of the source vector $\boldsymbol{j}$,
the cumulants can be obtained as derivatives evaluated at $\boldsymbol{j}=0$.
The derivation uses standard results from Gaussian integrals, matrix
calculus, and series expansions to express each cumulant in terms
of the parameters $A^{\alpha}$, $B^{\alpha}$, and $c^{\alpha}$.

The starting point for the calculations of the cumulants is the approximation
of the network as a quadratic mapping from latent space to data space
(Eq.~\ref{eq:quad_net_mapping}),

\begin{eqnarray}
x^{\alpha} & = & \frac{1}{2}\sum_{i,j}A_{ij}^{\alpha}z_{i}z_{j}+\sum_{i}B_{i}^{\alpha}z_{i}+c^{\alpha}\,.\label{eq:quadratic_approx}
\end{eqnarray}
Without loss of generality, we can always transform the latent variables
$z\sim\mathcal{N}(\mu,\Sigma)$ such that $\tilde{z}\sim\mathcal{N}(0,\mathbb{I})$.
This standardization simplifies the analytical treatment and allows
us to carry out all computations using standard Gaussian identities.
For each mixture component $k$, this transformation effectively replaces
the parameters as

\begin{eqnarray*}
\tilde{c}_{k}^{\alpha} & = & c^{\alpha}+B^{\alpha T}\mu_{k}+\frac{1}{2}\mu_{k}^{T}A^{\alpha}\mu_{k}\\
\tilde{B}_{k}^{\alpha} & = & \Sigma_{k}^{1/2}B^{\alpha}+\Sigma_{k}^{1/2}A^{\alpha}\mu_{k}\\
\tilde{A_{k}^{\alpha}} & = & \Sigma_{k}^{1/2}A^{\alpha}\Sigma_{k}^{1/2}\,.
\end{eqnarray*}
With a slight abuse of notation, we omit the tilde in this renaming
in the derivation for legibility, where the index $k$ stands for
the Gaussian component here.

We now use the expression above to compute the moment generating function,
averaging over the standard Gaussian distribution $z\propto\mathcal{N}(0,\mathbb{I})$
\begin{eqnarray*}
Z(\bm{j}) & = & \langle e^{\sum_{\alpha}j^{\alpha}x^{\alpha}}\rangle_{z}\\
 & = & \langle e^{\sum_{\alpha}j^{\alpha}c^{\alpha}+\sum_{\alpha,i}j^{\alpha}B_{i}^{\alpha}z_{i}+\frac{1}{2}\sum_{\alpha,i,j}j^{\alpha}A_{ij}^{\alpha}z_{i}z_{j}}\rangle_{z}\\
 & = & e^{\sum_{\alpha}j^{\alpha}c^{\alpha}}\int d^{n}z\frac{1}{(2\pi)^{n/2}}\bigg(e^{-\frac{1}{2}\sum_{i,j}z_{i}\delta_{ij}z_{j}}\times\\
 &  & \times e^{\sum_{\alpha}j^{\alpha}B_{i}^{\alpha}z_{i}+\frac{1}{2}\sum_{\alpha,i,j}j^{\alpha}A_{ij}^{\alpha}z_{i}z_{j}}\bigg)\,.
\end{eqnarray*}
We define a $\bm{j}-$dependent covariance matrix $\Sigma$ as
\begin{eqnarray*}
\Sigma(\bm{j}) & := & (\mathbb{I}-\sum_{\alpha}A^{\alpha}j^{\alpha})^{-1}\,.
\end{eqnarray*}
Thus, we get

\begin{eqnarray*}
Z(\bm{j}) & = & e^{\bm{j}^{T}\bm{c}}\int d^{n}z\frac{1}{(2\pi)^{n/2}}\\
 &  & \bigg(e^{-\frac{1}{2}(\bm{z}-\Sigma(\bm{j})B^{T}\bm{j})^{T}\Sigma^{-1}(\bm{j})(\bm{z}-\Sigma(\bm{j})B^{T}\bm{j})}\times\\
 &  & \times e^{\frac{1}{2}\bm{j^{T}}B\Sigma(\bm{j})B^{T}\bm{j}}\bigg)\\
 & = & e^{\bm{j}^{T}\bm{c}}e^{\frac{1}{2}\bm{j^{T}}B\Sigma(\bm{j})B^{T}\bm{j}}(\det(\Sigma(\bm{j})))^{1/2}\,.
\end{eqnarray*}
From here, the cumulant generating function directly follows as

\begin{eqnarray*}
W(\bm{j}) & = & \ln\,Z(\bm{j})\\
 & = & \bm{j}^{T}\bm{c}+\frac{1}{2}\bm{j^{T}}B(\mathbb{I}-\sum_{\alpha}A^{\alpha}j^{\alpha})^{-1}B^{T}\bm{j}\\
 &  & -\frac{1}{2}\ln\,\det(\mathbb{I}-\sum_{\alpha}A^{\alpha}j^{\alpha})\,.
\end{eqnarray*}
We expand the term $\Sigma(\bm{j})$ using the geometric series, given
that we will be only interested in derivatives by $j$ at $j=0$,
\begin{eqnarray*}
\Sigma(\bm{j}) & = & (\mathbb{I}-\sum_{\alpha}A^{\alpha}j^{\alpha})^{-1}\\
 & = & \sum_{k=0}^{\infty}\,(\sum_{\alpha}A^{\alpha}j^{\alpha})^{k}\,.
\end{eqnarray*}
To find a series expansion, we use that $\sum_{\alpha}A^{\alpha}j^{\alpha}$
is symmetric and therefore can be decomposed into its eigenvalues
$\lambda_{l}$. It thus follows
\begin{eqnarray*}
\ln\,\det(\mathbb{I}-\sum_{\alpha}A^{\alpha}j^{\alpha})) & = & \sum_{l}\,\ln(1-\lambda_{l})\\
 & = & -\sum_{l}\sum_{k=0}^{\infty}\frac{\lambda_{l}^{k}}{k}\\
 & = & -\sum_{k=0}^{\infty}\frac{\sum_{l}\lambda_{l}^{k}}{k}\\
 & = & -\sum_{k=1}^{\infty}\frac{\text{Tr}((\sum_{\alpha}A^{\alpha}j^{\alpha})^{k})}{k}\,.
\end{eqnarray*}
 We can now write the cumulant-generating function as an expansion
in $\bm{j}$,

\begin{eqnarray*}
W(\bm{j}) & = & \bm{j}^{T}\bm{c}+\frac{1}{2}\sum_{k=0}^{\infty}\bm{j^{T}}B(\sum_{\alpha}A^{\alpha}j^{\alpha})^{k}B^{T}\bm{j}\\
 &  & +\frac{1}{2}\sum_{k=1}^{\infty}\frac{\text{Tr}((\sum_{\alpha}A^{\alpha}j^{\alpha})^{k})}{k}\\
 & = & \sum_{k=1}^{\infty}\bigg[\delta_{k,1}\bm{j}^{T}\bm{c}\\
 &  & +(1-\delta_{k,1})\frac{1}{2}\bm{j^{T}}B(\sum_{\alpha}A^{\alpha}j^{\alpha})^{k-2}B^{T}\bm{j}\\
 &  & +\frac{1}{2}\frac{\text{Tr}((\sum_{\alpha}A^{\alpha}j^{\alpha})^{k})}{k}\bigg]\,.
\end{eqnarray*}
It is properly normalized as $W(\bm{0})=0.$

We can now read off the terms of different orders to obtain the cumulants,
which we do here for four orders. At first order we have

\begin{eqnarray*}
W^{1}(\bm{j}) & = & \bm{j}^{T}\bm{c}+\frac{1}{2}\text{Tr}(\sum_{\alpha}A^{\alpha}j^{\alpha})\\
\langle\langle x^{\alpha}\rangle\rangle & = & c^{\alpha}+\frac{1}{2}\text{Tr}(A^{\alpha})
\end{eqnarray*}
which yields the mean. At second order, we get the covariance

\begin{eqnarray*}
W^{2}(\bm{j}) & = & \frac{1}{2}\bm{j^{T}}BB^{T}\bm{j}+\frac{1}{4}\text{Tr}((\sum_{\alpha}A^{\alpha}j^{\alpha})^{2})\\
\langle\langle x^{\alpha}x^{\beta}\rangle\rangle & = & [BB^{T}]^{\alpha\beta}+\frac{1}{4}\frac{\partial^{2}}{\partial j_{\alpha}\partial j_{\beta}}\text{Tr}(\sum_{\alpha^{\prime},\beta^{\prime}}(A^{\alpha^{\prime}}j^{\alpha^{\prime}})(A^{\beta^{\prime}}j^{\beta^{\prime}}))\\
 & = & [BB^{T}]^{\alpha\beta}+\frac{1}{4}\sum_{(\alpha^{\prime},\beta^{\prime})\in\mathcal{P}(\alpha,\beta)}\text{Tr}(A^{\alpha^{\prime}}A^{\beta^{\prime}})\,,
\end{eqnarray*}
where $\mathcal{P}(\alpha,\ldots,\beta)$ is the set of all permutations
of the given set of indices. At third order

\begin{eqnarray*}
W^{3}(\bm{j}) & = & \frac{1}{2}\bm{j^{T}}B(\sum_{\alpha}A^{\alpha}j^{\alpha})B^{T}\bm{j}\\
 &  & +\frac{1}{6}\text{Tr}((\sum_{\alpha}A^{\alpha}j^{\alpha})^{3})
\end{eqnarray*}
the third order cumulants follow as

\begin{eqnarray*}
\langle\langle x^{\alpha}x^{\beta}x^{\gamma}\rangle\rangle & = & \frac{1}{2}\sum_{(\alpha^{\prime},\beta^{\prime},\gamma^{\prime})\in\mathcal{P}(\alpha,\beta,\gamma)}[BA^{\beta^{\prime}}B^{T}]^{\alpha^{\prime}\gamma^{\prime}}\\
 &  & +\frac{1}{6}\sum_{(\alpha^{\prime},\beta^{\prime},\gamma^{\prime})\in\mathcal{P}(\alpha,\beta,\gamma)}\text{Tr}(A^{\alpha^{\prime}}A^{\beta^{\prime}}A^{\gamma^{\prime}})\,.
\end{eqnarray*}
At fourth order

\begin{eqnarray*}
W^{4}(\bm{j}) & = & \frac{1}{2}\bm{j^{T}}B(\sum_{\alpha}A^{\alpha}j^{\alpha})^{2}B^{T}\bm{j}\\
 &  & +\frac{1}{8}\text{Tr}((\sum_{\alpha}A^{\alpha}j^{\alpha})^{4})
\end{eqnarray*}
we obtain the fourth order cumulants as

\begin{eqnarray*}
 &  & \langle\langle x^{\alpha}x^{\beta}x^{\gamma}x^{\delta}\rangle\rangle\\
 & = & \frac{1}{2}\sum_{(\alpha^{\prime},\beta^{\prime},\gamma^{\prime},\delta^{\prime})\in\mathcal{P}(\alpha,\beta,\gamma,\delta)}[BA^{\alpha^{\prime}}A^{\beta^{\prime}}B^{T}]^{\gamma^{\prime}\delta^{\prime}}\\
 &  & +\frac{1}{8}\sum_{(\alpha^{\prime},\beta^{\prime},\gamma^{\prime},\delta^{\prime})\in\mathcal{P}(\alpha,\beta,\gamma,\delta)}\text{Tr}(A^{\alpha^{\prime}}A^{\beta^{\prime}}A^{\gamma^{\prime}}A^{\delta^{\prime}})\,.
\end{eqnarray*}
In Fig.~\ref{fig:cumulants}, we show results for the covariance
$\langle\langle x^{\alpha}x^{\beta}\rangle\rangle$, the third order
cumulants $\langle\langle x^{\alpha}x^{\beta}x^{\beta}\rangle\rangle$,
and the fourth order cumulants $\langle\langle x^{\alpha}x^{\alpha}x^{\beta}x^{\beta}\rangle\rangle$.

\section{Curvature\protect\label{sec:Sup-Curvature}}

In this section, we derive the key geometric quantities that describe
the curvature of data manifolds when mapped from a flat latent space.
Specifically, we will focus on the Riemannian curvature tensor ${R_{klj}}^{i}$,
the sectional curvature $K(e_{i},e_{j})$ and the scalar curvature
$R$.

We start by considering the mapping of latent variables $z_{j}$\LyXZeroWidthSpace{}
to data space $x^{\alpha}$ through a quadratic function given by
Eq.~\ref{eq:quad_net_mapping}). For simplicity, we adopt the Einstein
summation convention throughout this derivation. Greek letters denote
indices in data space (upper indices), while Latin letters refer to
indices in latent space (lower indices).

\paragraph*{Step 1: Tangent Vectors}

To understand how latent variables act as coordinates of the data
space, we first compute tangent vectors (see \citep{Lee2012introduction}~Ch.3).
These vectors represent partial derivatives of data coordinates with
respect to latent variables:

\begin{eqnarray}
e_{i}^{\alpha} & = & \frac{\partial x^{\alpha}}{\partial z_{i}}=B_{i}^{\alpha}+A_{ij}^{\alpha}z_{j}\,.\label{eq:tangent_vectors_app}
\end{eqnarray}
This equation shows how infinitesimal changes in each latent variable
influence the position within the manifold.

\paragraph{Step 2: Metric Tensor}

The metric tensor encodes local geometric information such as distances
and angles between directions on the manifold. It is constructed using
the tangent vectors (see \citep{Lee2018introduction}~Ch.2):

\begin{eqnarray*}
g_{ij} & = & e_{i}^{\alpha}e_{j}^{\alpha}\,.
\end{eqnarray*}
Expressed in terms of the tangent vectors (Eq.~\ref{eq:tangent_vectors_app})
the metric tensor is

\begin{eqnarray*}
g_{ij} & = & (B_{i}^{\alpha}+A_{ik}^{\alpha}z_{k})(B_{j}^{\alpha}+A_{jl}^{\alpha}z_{l})\\
 & = & \underbrace{B_{i}^{\alpha}B_{j}^{\alpha}}_{:=\gamma_{ij}}+\underbrace{(B_{i}^{\alpha}A_{jk}^{\alpha}+B_{j}^{\alpha}A_{ik}^{\alpha})}_{:=w_{ij,k}}z_{k}+\underbrace{A_{ik}^{\alpha}A_{jl}^{\alpha}}_{:=h_{ik,jl}}z_{k}z_{l}\\
 & = & \gamma_{ij}+w_{ij,k}z_{k}+h_{ik,jl}z_{k}z_{l}\,,
\end{eqnarray*}
where we introduce $\gamma_{ij}$\LyXZeroWidthSpace , $w_{ij,k}$\LyXZeroWidthSpace ,
and $h_{ik,jl}$\LyXZeroWidthSpace{} as the coefficients for the different
powers of $z$ of the metric tensor.

For the derivation of the Riemannian curvature, we will also need
the partial derivatives of the metric tensor:

\begin{eqnarray*}
g_{ij,k} & = & \frac{\partial}{\partial z_{k}}g_{ij}\\
 & = & \frac{\partial}{\partial z_{k}}(\gamma_{ij}+w_{ij,m}z_{km}+h_{ij,ml}z_{m}z_{l})\\
 & = & w_{ij,k}+(h_{ik,jl}+h_{il,jk})z_{l}\,.
\end{eqnarray*}

\paragraph*{Note on Taylor expansion around $z=0$}

To simplify our calculations, we utilize a Taylor expansion around
$z=0$. This approach is chosen without loss of generality because
the expansion could equivalently be performed around any point $z=z^{\ast}$.
By focusing on $z=0$, we can derive expressions for the curvature
tensors that are valid at this reference point. This expression can
then be generalized to other points to obtain $R_{jkl}^{i}(z=z^{\ast})$.

\paragraph{Step 3: Inverse Metric Tensor}

Next, we determine the inverse metric tensor (where we also use upper
indices) by expanding it around $z=0$:

\begin{eqnarray}
g^{ij} & = & g_{(0)}^{ij}+g_{(1),k}^{ij}z_{k}+g_{(2),kl}^{ij}z_{k}z_{l}+\mathcal{O}(z^{3})\,.\label{eq:inverse-metric-tensor}
\end{eqnarray}
These components are calculated by ensuring that their product with
the metric tensor $g_{jm}$\LyXZeroWidthSpace{} equals the identity
matrix (see \citep{Lee2012introduction}~Ch.8):

\begin{eqnarray*}
g^{ij}g_{jm} & = & \delta_{im}\\
 & = & (g_{(0)}^{ij}+g_{(1),k}^{ij}z_{k}+g_{(2),kl}^{ij}z_{k}z_{l}+\mathcal{O}(z^{3}))\times\\
 &  & \times(\gamma_{jm}+w_{jm,k}z_{k}+h_{jk,ml}z_{k}z_{l})\\
 & = & g_{(0)}^{ij}\gamma_{jm}+(g_{(0)}^{ij}w_{jm,k}+g_{(1),k}^{ij}\gamma_{jm})z_{k}\\
 &  & +(g_{(0)}^{ij}h_{jk,ml}+g_{(1),k}^{ij}w_{jm,l}+g_{(2),kl}^{ij}\gamma_{jm})z_{k}z_{l}\\
 &  & +\mathcal{O}(z^{3})\,.
\end{eqnarray*}
Since the inverse relationship must hold for all $z$ and the polynomial
of $z$ of different powers are independent, the coefficients in front
of each power of $z$ must vanish separately. We can hence read of
the terms up to quadratic order. At zeroth order in $z$ we have

\begin{eqnarray}
g_{(0)}^{ij} & = & (\gamma^{-1})_{ij}=\gamma^{ij}\,.\label{eq:zeroth_inv_tensor}
\end{eqnarray}
The first order term $\propto z$ yields

\begin{eqnarray}
0 & = & g_{(0)}^{ij}w_{jm,k}+g_{(1),k}^{ij}\gamma_{jm}\nonumber \\
g_{(1),k}^{ij} & = & -g_{(0)}^{in}w_{nm,k}g_{(0)}^{mj}=-\gamma^{in}w_{nm,k}\gamma^{mj}\,,\label{eq:first_inv_tensor}
\end{eqnarray}
where we used Eq.~\eqref{eq:zeroth_inv_tensor}. The second order
yield

\begin{eqnarray*}
0 & = & g_{(0)}^{ij}h_{jk,ml}+g_{(1),k}^{ij}w_{jm,l}+g_{(2),kl}^{ij}\gamma_{jm}\\
g_{(2),kl}^{ij} & = & -g_{(0)}^{in}h_{nk,ml}g_{(0)}^{mj}-g_{(1),k}^{in}w_{nm,l}g_{(0)}^{mj}\\
 & \stackrel{(\ref{eq:zeroth_inv_tensor}\&\ref{eq:first_inv_tensor})}{=} & -\gamma^{in}h_{nk,ml}\gamma^{mj}+\gamma^{io}w_{om,k}\gamma^{mn}w_{nm,l}\gamma^{mj}\,.
\end{eqnarray*}

\paragraph*{Step 4: Christoffel Symbols}

Christoffel symbols are essential for computing curvature because
they capture connections between tangent spaces (see \citep{Lee2018introduction}~Ch.5).
We expand them up to second order in $z$:

\textbf{
\begin{eqnarray*}
\Gamma_{kl}^{i} & = & \frac{1}{2}g^{im}(g_{mk,l}+g_{ml,k}-g_{kl,m})\\
 & = & \frac{1}{2}g^{im}(w_{mk,l}+w_{ml,k}-w_{kl,m})\\
 &  & +\frac{1}{2}g^{im}((h_{ml,kn}+h_{mn,kl}+h_{mk,ln}+h_{mn,lk})z_{n})\\
 &  & -\frac{1}{2}g^{im}((h_{km,ln}+h_{kn,lm})z_{n})\\
 & = & \frac{1}{2}g^{im}(w_{mk,l}+w_{ml,k}-w_{kl,m})+g^{im}h_{mn,kl}z_{n}\\
 & \overset{\eqref{eq:inverse-metric-tensor}}{=} & \frac{1}{2}(\gamma^{im}-\gamma^{in}w_{no,a}\gamma^{om}z_{a}+\mathcal{O}(z^{2}))\times\\
 &  & \times(w_{mk,l}+w_{ml,k}-w_{kl,m})\\
 &  & +(\gamma^{im}+\mathcal{O}(z))h_{mn,kl}z_{n}\\
 & = & (\gamma^{im}-\gamma^{in}w_{no,a}\gamma^{om}z_{a})B_{m}^{\alpha}A_{kl}^{\alpha}\\
 &  & +\gamma^{im}h_{mn,kl}z_{n}+\mathcal{O}(z^{2})\,.
\end{eqnarray*}
}For the last step, we used that \textbf{
\begin{eqnarray*}
w_{mk,l}+w_{ml,k}-w_{kl,m} & = & B_{m}^{\alpha}A_{kl}^{\alpha}+B_{k}^{\alpha}A_{ml}^{\alpha}+B_{m}^{\alpha}A_{lk}^{\alpha}\\
 &  & +B_{l}^{\alpha}A_{mk}^{\alpha}-B_{k}^{\alpha}A_{lm}^{\alpha}-B_{l}^{\alpha}A_{km}^{\alpha}\,.\\
 & = & 2B_{m}^{\alpha}A_{kl}^{\alpha}
\end{eqnarray*}
}The partial derivative of the Christoffel symbols are also required,
which up to linear order are:

\begin{eqnarray*}
\frac{\partial\Gamma_{lj}^{i}}{\partial z^{k}} & = & -\gamma^{in}w_{no,k}\gamma^{om}B_{m}^{\alpha}A_{lj}^{\alpha}+\gamma^{im}h_{mk,lj}+\mathcal{O}(z)\,.
\end{eqnarray*}

\paragraph*{Step 5: Riemannian Curvature Tensor}

With all necessary components, we can now compute the Riemannian curvature
tensor which quantifies intrinsic bending or warping at a point on
manifold (see \citep{Lee2018introduction}~Ch.7):

\begin{eqnarray*}
{R_{klj}}^{i} & = & \frac{\partial\Gamma_{lj}^{i}}{\partial z^{k}}-\frac{\partial\Gamma_{kj}^{i}}{\partial z^{l}}+\big(\Gamma_{kp}^{i}\Gamma_{lj}^{p}-\Gamma_{lp}^{i}\Gamma_{kj}^{p}\big)\,.
\end{eqnarray*}
We perform the different parts of this calculation

\begin{eqnarray*}
 &  & \frac{\partial\Gamma_{lj}^{i}}{\partial z^{k}}-\frac{\partial\Gamma_{kj}^{i}}{\partial z^{l}}\\
 & = & -\gamma^{in}w_{no,k}\gamma^{om}B_{m}^{\alpha}A_{lj}^{\alpha}\\
 &  & +\gamma^{in}w_{no,l}\gamma^{om}B_{m}^{\alpha}A_{kj}^{\alpha}\\
 &  & +\gamma^{im}h_{mk,lj}-\gamma^{im}h_{ml,kj}+\mathcal{O}(z)\\
 & = & \gamma^{in}B_{m}^{\alpha}\gamma^{om}B_{n}^{\beta}(-A_{ok}^{\beta}A_{lj}^{\alpha}+A_{ol}^{\beta}A_{kj}^{\alpha})\\
 &  & +\gamma^{in}B_{m}^{\alpha}\gamma^{om}B_{o}^{\beta}(-A_{nk}^{\beta}A_{lj}^{\alpha}+A_{nl}^{\beta}A_{kj}^{\alpha})\\
 &  & +\gamma^{im}h_{mk,lj}-\gamma^{im}h_{ml,kj}+\mathcal{O}(z)\,.
\end{eqnarray*}
The product of the Christoffel symbols becomes:

\begin{eqnarray*}
\Gamma_{kp}^{i}\Gamma_{lj}^{p} & = & \gamma^{im}B_{m}^{\alpha}A_{kp}^{\alpha}\gamma^{pn}B_{n}^{\beta}A_{lj}^{\beta}+\mathcal{O}(z)\,.
\end{eqnarray*}
Further, the substraction of products of Christoffel symbols is:

\begin{eqnarray*}
 &  & \Gamma_{kp}^{i}\Gamma_{lj}^{p}-\Gamma_{lp}^{i}\Gamma_{kj}^{p}\\
 & = & \gamma^{im}B_{m}^{\alpha}A_{kp}^{\alpha}\gamma^{pn}B_{n}^{\beta}A_{lj}^{\beta}\\
 &  & -\gamma^{im}B_{m}^{\alpha}A_{lp}^{\alpha}\gamma^{pn}B_{n}^{\beta}A_{kj}^{\beta}+\mathcal{O}(z)\\
 & = & \gamma^{im}B_{m}^{\alpha}\gamma^{pn}B_{n}^{\beta}(A_{kp}^{\alpha}A_{lj}^{\beta}-A_{lp}^{\alpha}A_{kj}^{\beta})+\mathcal{O}(z)\,.
\end{eqnarray*}
The Riemannian curvature tensor follows thus as:

\begin{eqnarray*}
{R_{klj}}^{i} & = & \frac{\partial\Gamma_{lj}^{i}}{\partial z^{k}}-\frac{\partial\Gamma_{kj}^{i}}{\partial z^{l}}+\big(\Gamma_{kp}^{i}\Gamma_{lj}^{p}-\Gamma_{lp}^{i}\Gamma_{kj}^{p}\big)\\
 & = & \gamma^{in}B_{m}^{\alpha}\gamma^{pm}B_{n}^{\beta}(-A_{pk}^{\beta}A_{lj}^{\alpha}+A_{pl}^{\beta}A_{kj}^{\alpha})\\
 &  & +\gamma^{in}B_{m}^{\alpha}\gamma^{pm}B_{p}^{\beta}(-A_{nk}^{\beta}A_{lj}^{\alpha}+A_{nl}^{\beta}A_{kj}^{\alpha})\\
 &  & +\gamma^{im}(A_{mk}^{\alpha}A_{lj}^{\alpha}-A_{ml}^{\alpha}A_{kj}^{\alpha})\\
 &  & +\gamma^{in}B_{m}^{\beta}\gamma^{pm}B_{n}^{\alpha}(A_{kp}^{\alpha}A_{lj}^{\beta}-A_{lp}^{\alpha}A_{kj}^{\beta})+\mathcal{O}(z)\\
 & = & \gamma^{in}B_{m}^{\alpha}\gamma^{pm}B_{p}^{\beta}(-A_{nk}^{\beta}A_{lj}^{\alpha}+A_{nl}^{\beta}A_{kj}^{\alpha})\\
 &  & +\gamma^{im}(A_{mk}^{\alpha}A_{lj}^{\alpha}-A_{ml}^{\alpha}A_{kj}^{\alpha})+\mathcal{O}(z)\\
 & = & \gamma^{im}B_{n}^{\alpha}\gamma^{pn}B_{p}^{\beta}(-A_{mk}^{\beta}A_{lj}^{\alpha}+A_{ml}^{\beta}A_{kj}^{\alpha})\\
 &  & +\gamma^{im}(A_{mk}^{\alpha}A_{lj}^{\alpha}-A_{ml}^{\alpha}A_{kj}^{\alpha})+\mathcal{O}(z)\\
 & = & \gamma^{im}A_{lj}^{\alpha}(A_{mk}^{\alpha}-B_{n}^{\alpha}\gamma^{pn}B_{p}^{\beta}A_{mk}^{\beta})\\
 &  & -\gamma^{im}A_{kj}^{\alpha}(A_{ml}^{\alpha}-B_{n}^{\alpha}\gamma^{pn}B_{p}^{\beta}A_{ml}^{\beta})+\mathcal{O}(z)\\
 & = & \gamma^{im}A_{jl}^{\alpha}(\delta_{\alpha\beta}-B_{n}^{\alpha}\gamma^{pn}B_{p}^{\beta})A_{mk}^{\beta}\\
 &  & -\gamma^{im}A_{kj}^{\alpha}(\delta_{\alpha\beta}-B_{n}^{\alpha}\gamma^{pn}B_{p}^{\beta})A_{ml}^{\beta}+\mathcal{O}(z)\\
 & = & \gamma^{im}(\delta_{\alpha\beta}-B_{n}^{\alpha}\gamma^{pn}B_{p}^{\beta})(A_{jl}^{\alpha}A_{mk}^{\beta}-A_{kj}^{\alpha}A_{ml}^{\beta})\\
 &  & +\mathcal{O}(z)\,.
\end{eqnarray*}
For $z=0$, it follows that: 

\begin{eqnarray*}
 &  & {R_{klj}}^{i}(z=0)\\
 & = & \gamma^{im}(\delta_{\alpha\beta}-B_{n}^{\alpha}\gamma^{pn}B_{p}^{\beta})(A_{jl}^{\alpha}A_{mk}^{\beta}-A_{kj}^{\alpha}A_{ml}^{\beta})\\
 & = & g^{im}(z=0)(\delta_{\alpha\beta}-e_{n}^{\alpha}(z=0)g^{pn}(z=0)e_{p}^{\beta}(z=0))\times\\
 &  & \times(A_{jl}^{\alpha}A_{mk}^{\beta}-A_{kj}^{\alpha}A_{ml}^{\beta})\,.
\end{eqnarray*}
Given that we performed the calculations here around $z=0$ without
loss of generality, we can conclude for all $z$ that 

\begin{eqnarray*}
{R_{klj}}^{i}(z) & = & g^{im}(z)\underbrace{(\delta_{\alpha\beta}-e_{n}^{\alpha}(z)g^{pn}(z)e_{p}^{\beta}(z))}_{:=P^{\alpha\beta}}\times\\
 &  & \times(A_{jl}^{\alpha}A_{mk}^{\beta}-A_{kj}^{\alpha}A_{ml}^{\beta})\,.
\end{eqnarray*}
Note that there is no $z$-dependency for the second derivative of
Eq.~\eqref{eq:quad_net_mapping}.

Further, we make use of the fact that $P^{\alpha\beta}=P^{\alpha\gamma}P^{\gamma\beta}$,
to introduce 
\begin{eqnarray*}
H_{ij}^{\gamma} & : & =P^{\alpha\gamma}A_{jl}^{\alpha}\,.
\end{eqnarray*}
With this definition, we can write the curvature tensor as 

\begin{eqnarray*}
{R_{klj}}^{i} & = & g^{im}(H_{jl}^{\gamma}H_{mk}^{\gamma}-H_{kj}^{\gamma}H_{ml}^{\gamma})\,.
\end{eqnarray*}

Like this we recover the description used in \citep{Jones2024manifold}.

After simplifying terms using the projection operator $P^{\alpha\beta}$,
it becomes:

\begin{eqnarray}
{R_{klj}}^{i} & = & g^{im}P^{\alpha\beta}(A_{jl}^{\alpha}A_{mk}^{\beta}-A_{kj}^{\alpha}A_{ml}^{\beta}).\label{eq:app-riemannian-curvature}
\end{eqnarray}
This expression captures interactions between different dimensions
contributing towards the overall shape distortion.

\paragraph*{Step 6: Sectional Curvature}

Sectional curvature measures how much the manifold bends along specific
two-dimensional planes spanned by pairs of tangent vectors (see \citep{Lee2018introduction}~Ch.7):

\begin{eqnarray*}
K(e_{i},e_{j}) & = & \frac{\langle R(e_{i},e_{j})e_{i},e_{j}\rangle}{\langle e_{i},e_{i}\rangle\langle e_{j},e_{j}\rangle-\langle e_{i},e_{j}\rangle^{2}}\\
 & = & \frac{g_{jm}{R_{jii}}^{m}}{g_{ii}g_{jj}-g_{ij}^{2}}\,.
\end{eqnarray*}
We enter Eq.~\ref{eq:app-riemannian-curvature} into this formula:

\begin{eqnarray*}
K(e_{i},e_{j}) & = & \frac{P^{\alpha\beta}(A_{ii}^{\alpha}A_{jj}^{\beta}-A_{ij}^{\alpha}A_{ij}^{\beta})}{g_{ii}g_{jj}-g_{ij}^{2}}\,.
\end{eqnarray*}

\paragraph*{Step 7: Scalar Curvature}

Finally calculating the scalar curvature provides an aggregate measure
summarizing total intrinsic curvature occurring throughout any direction
at a given location on the manifold.

To calculate this, we first determine the Ricci curvature (see \citep{Lee2018introduction}~Ch.7): 

\begin{eqnarray*}
R_{jl} & = & {R_{cjl}}^{c}\\
 & = & g^{cm}P^{\alpha\beta}(A_{jl}^{\alpha}A_{mc}^{\beta}-A_{cj}^{\alpha}A_{ml}^{\beta})\,.
\end{eqnarray*}
Finally, we arrive at the scalar curvature (see \citep{Lee2018introduction}~Ch.7):

\begin{eqnarray*}
R & = & g^{jl}R_{jl}\\
 & = & g^{jl}g^{cm}P^{\alpha\beta}(A_{jl}^{\alpha}A_{mc}^{\beta}-A_{cj}^{\alpha}A_{ml}^{\beta})\,.
\end{eqnarray*}

\begin{widetext}

\section{Supplementary figures and tables\protect\label{sec:Supplementary-Figures}}

In the last section, we show that our results are consistent across
recording sessions. In order to show that, we present the results
obtained for session L\_RS\_250717 also for the two sessions L\_RS\_090817
and L\_RS\_100817.

\begin{figure}[H]
\includegraphics{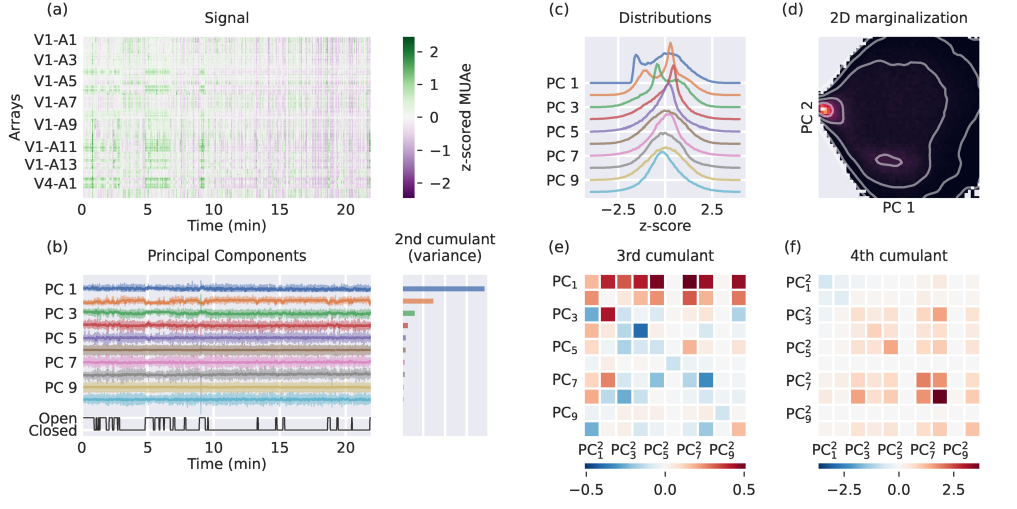}\caption{\protect\label{fig:data-overview-1}\textbf{Neuronal data exhibit
higher-order correlations. (a) }The original signal from one recording
session (L\_RS\_090817), sampled at 100~Hz over 22.0~minutes. The
signal is a multi-unit activity envelope (MUAe) containing 774 channels
from 16 recording electrode arrays (14 in V1 and 2 in V4). Each channel
is independently $z$-scored (zero mean, unit variance) to ensure
consistent scaling across neurons. \textbf{(b) }Time evolution of
the first ten PCA variables (PCs) of the $z$-scored data. Transparent
curves represent raw PCs, while solid curves show temporal smoothing
using a Gaussian kernel with 1~s width. The labels of the behavioral
state (eyes open or eyes closed) are shown below. The rapid decay
of the eigenvalues $\sigma_{i}^{2}$ of the covariance matrix (2nd-order
cumulants; on the right), corresponding to the shown PCs, is an indicator
that the neural data lie on a lower-dimensional manifold. \textbf{(c)
}Marginal distribution of the first ten PCs. While some appear Gaussian,
others clearly deviate from Gaussianity. \textbf{(d)} Two-dimensional
marginalization along the first and second PCs. \textbf{(e-f)} 3\protect\textsuperscript{rd}
and 4\protect\textsuperscript{th}~order cumulant. For both cases,
the signals along the PCs are first divided by their standard deviation
$\sigma_{i}$ to normalize for the amplitude. Thus, panel (e) portrays
$x_{i}/\sigma_{i}$ against $x_{j}^{2}/\sigma_{j}^{2}$ and panel
(f) $x_{i}^{2}/\sigma_{i}^{2}$ against $x_{j}^{2}/\sigma_{j}^{2}$.
Color bars indicate cumulant magnitudes.}
\end{figure}

\begin{figure}[H]
\includegraphics{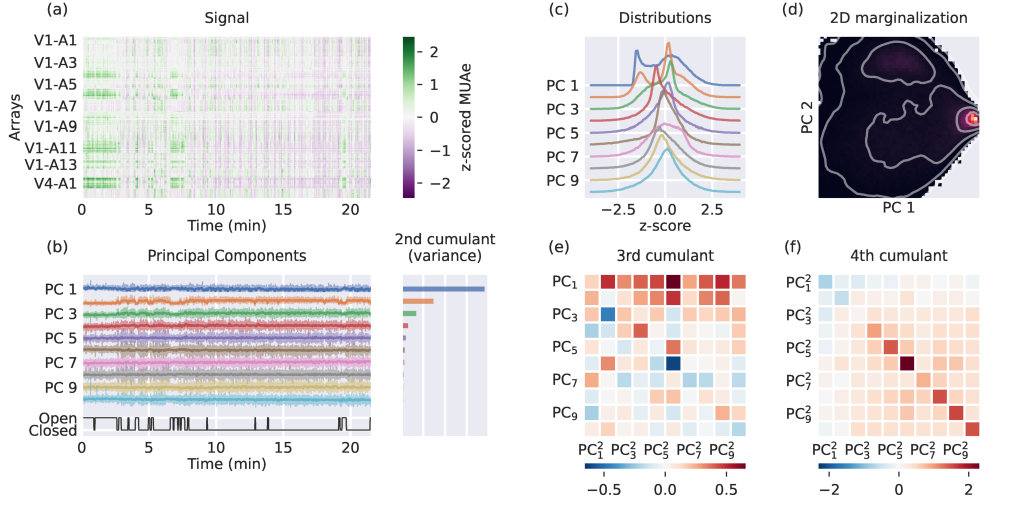}\caption{\protect\label{fig:data-overview-2}\textbf{Neuronal data exhibit
higher-order correlations. (a) }The original signal from one recording
session (L\_RS\_100817), sampled at 100~Hz over 21.6~minutes. The
signal is a multi-unit activity envelope (MUAe) containing 838 channels
from 16 recording electrode arrays (14 in V1 and 2 in V4). Each channel
is independently $z$-scored (zero mean, unit variance) to ensure
consistent scaling across neurons. \textbf{(b) }Time evolution of
the first ten PCA variables (PCs) of the $z$-scored data. Transparent
curves represent raw PCs, while solid curves show temporal smoothing
using a Gaussian kernel with 1~s width. The labels of the behavioral
state (eyes open or eyes closed) are shown below. The rapid decay
of the eigenvalues $\sigma_{i}^{2}$ of the covariance matrix (2nd-order
cumulants; on the right), corresponding to the shown PCs, is an indicator
that the neural data lie on a lower-dimensional manifold. \textbf{(c)
}Marginal distribution of the first ten PCs. While some appear Gaussian,
others clearly deviate from Gaussianity. \textbf{(d)} Two-dimensional
marginalization along the first and second PCs. \textbf{(e-f)} 3\protect\textsuperscript{rd}
and 4\protect\textsuperscript{th}~order cumulant. For both cases,
the signals along the PCs are first divided by their standard deviation
$\sigma_{i}$ to normalize for the amplitude. Thus, panel (e) portrays
$x_{i}/\sigma_{i}$ against $x_{j}^{2}/\sigma_{j}^{2}$ and panel
(f) $x_{i}^{2}/\sigma_{i}^{2}$ against $x_{j}^{2}/\sigma_{j}^{2}$.
Color bars indicate cumulant magnitudes.}
\end{figure}
\begin{figure}[H]
\includegraphics{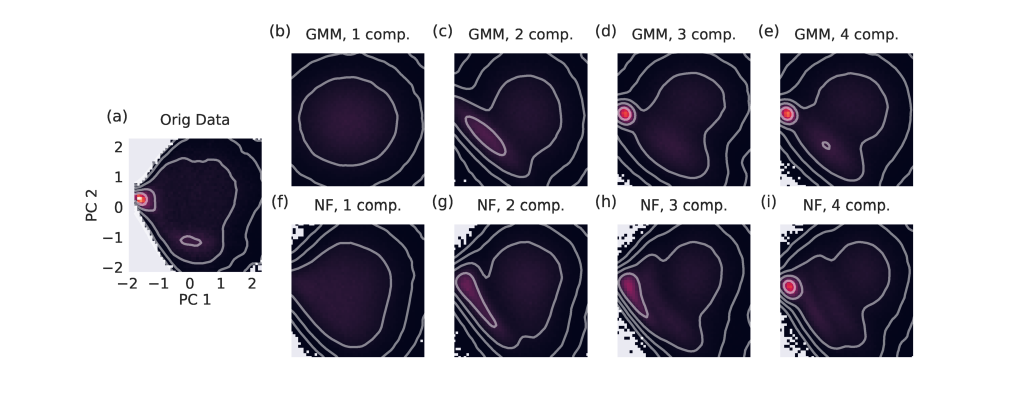}\caption{\protect\label{fig:component-comparison-1}\textbf{Effect of number
of latent components on the approximation capabilities: Marginalized
distributions} for the experimental data and learned models. \textbf{(a)
}Marginalized distribution of one session of experimental data (L\_RS\_090817)
along the first and second PCA directions. Light-gray isolines represent
equal likelihood. \textbf{(b--e)} Marginalized probability densities
for data sampled from Gaussian Mixture Models (GMMs) with increasing
numbers of latent components (1 to 4). \textbf{(f--i)} Corresponding
distributions for Normalizing Flows (NFs), also with 1 to 4 latent
components.}
\end{figure}

\begin{figure}[H]
\includegraphics{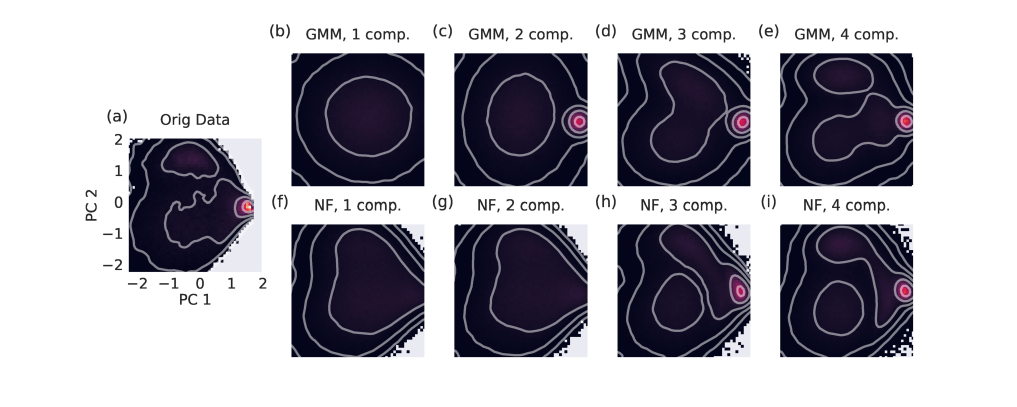}\caption{\protect\label{fig:component-comparison-2}\textbf{Effect of number
of latent components on the approximation capabilities: Marginalized
distributions} for the experimental data and learned models. \textbf{(a)
}Marginalized distribution of one session of experimental data (L\_RS\_100817)
along the first and second PCA directions. Light-gray isolines represent
equal likelihood. \textbf{(b--e)} Marginalized probability densities
for data sampled from Gaussian Mixture Models (GMMs) with increasing
numbers of latent components (1 to 4). \textbf{(f--i)} Corresponding
distributions for Normalizing Flows (NFs), also with 1 to 4 latent
components.}
\end{figure}

\end{widetext}

\begin{figure}[H]
\centering
\includegraphics{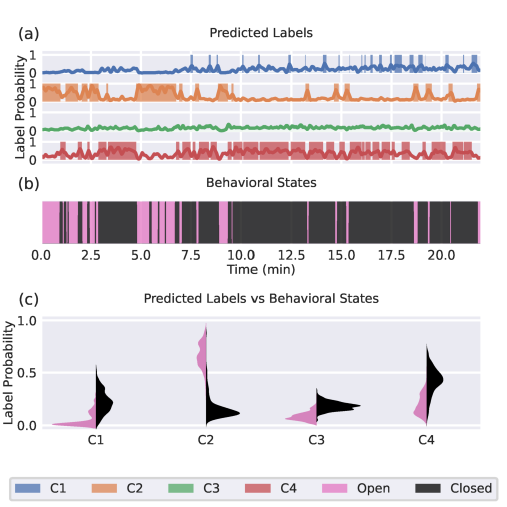}\caption{\protect\label{fig:label-alignment-1}\textbf{Distinct mixture components
in the latent space represent behavioral states (session L\_RS\_090817).
(a)} For each of the four components (C1--C4), smoothed (Gaussian
filter, $\sigma=3\,\text{s}$), normalized likelihoods are shown as
solid curves. Labels are assigned by selecting the component with
the largest label probability at each time point, indicated by background
shading. \textbf{(b)} Ground-truth behavioral states -- eyes open
(pink) and eyes closed (black) -- are shown for direct comparison
with the labels predicted in (a). \textbf{(c)} The label probabilities
are compared against the behavioral states. For each of the components,
smoothed histograms display the distributions of label probabilities
for the two behavioral states (eyes open: left, eyes closed: right).}
\end{figure}

\begin{figure}[H]
\centering
\includegraphics{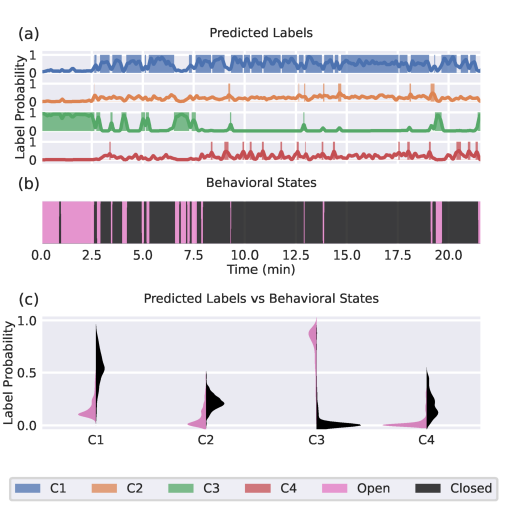}\caption{\protect\label{fig:label-alignment-2}\textbf{Distinct mixture components
in the latent space represent behavioral states (session L\_RS\_100817).
(a)} For each of the four components (C1--C4), smoothed (Gaussian
filter, $\sigma=3\,\text{s}$), normalized likelihoods are shown as
solid curves. Labels are assigned by selecting the component with
the largest label probability at each time point, indicated by background
shading. \textbf{(b)} Ground-truth behavioral states -- eyes open
(pink) and eyes closed (black) -- are shown for direct comparison
with the labels predicted in (a). \textbf{(c)} The label probabilities
are compared against the behavioral states. For each of the components,
smoothed histograms display the distributions of label probabilities
for the two behavioral states (eyes open: left, eyes closed: right).}
\end{figure}

\begin{widetext}

\begin{figure}[H]
\includegraphics{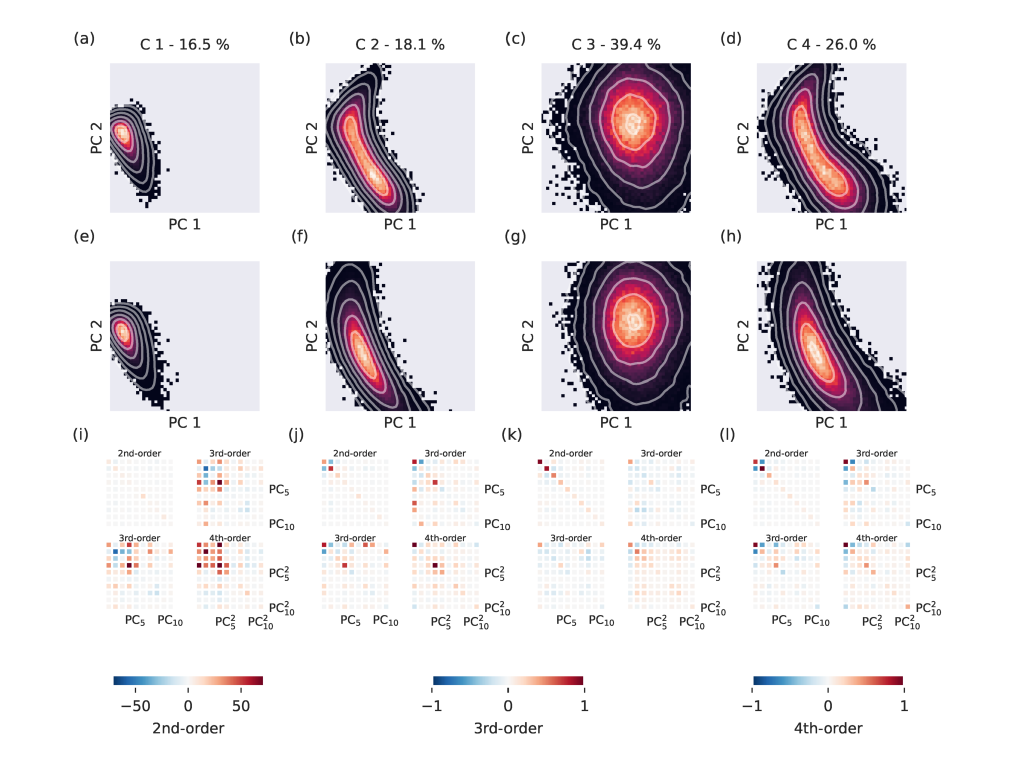}\caption{\protect\label{fig:cumulants-1}\textbf{Statistical characteristics
of the latent components: Approximate marginalized distributions}
for experimental data and learned models. \textbf{(a-d)} Marginalized
distributions of each latent component from the experimental session
(L\_RS\_090817), projected onto the first two PCA directions. Light-gray
contours indicate isolines of equal likelihood. \textbf{(e-h)} Marginalized
distribution of each component when the inverse mapping of the network
is approximated by a quadratic function (Eq.~\ref{eq:quad_net_mapping}).
\textbf{(i-l)} 2\protect\textsuperscript{nd} to 4\protect\textsuperscript{th}~order
cumulants computed from this approximation.}
\end{figure}

\begin{figure}[H]
\includegraphics{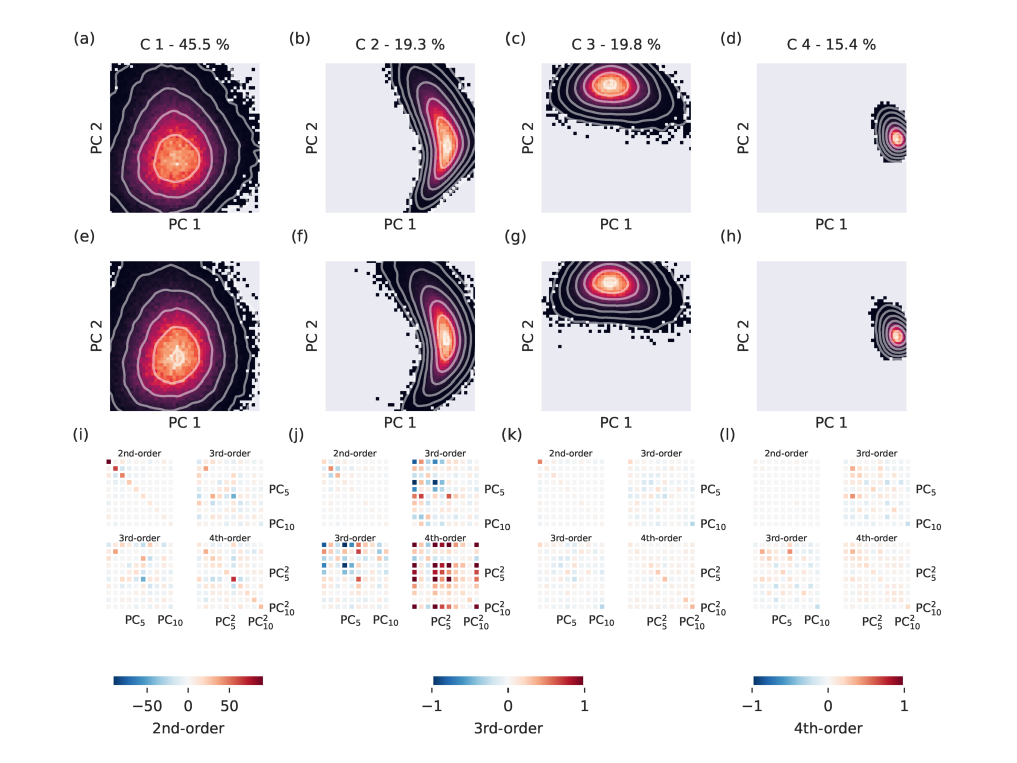}\caption{\protect\label{fig:cumulants-2}\textbf{Statistical characteristics
of the latent components: Approximate marginalized distributions}
for experimental data and learned models. \textbf{(a-d)} Marginalized
distributions of each latent component from the experimental session
(L\_RS\_100817), projected onto the first two PCA directions. Light-gray
contours indicate isolines of equal likelihood. \textbf{(e-h)} Marginalized
distribution of each component when the inverse mapping of the network
is approximated by a quadratic function (Eq.~\ref{eq:quad_net_mapping}).
\textbf{(i-l)} 2\protect\textsuperscript{nd} to 4\protect\textsuperscript{th}~order
cumulants computed from this approximation.}
\end{figure}

\begin{table}[H]
\centering
\begin{tabular}{|c|c|}
\hline 
 & Sectional curvature\tabularnewline
\hline 
\hline 
Component 1 & $-0.048$\tabularnewline
\hline 
Component 2 & $-0.035$\tabularnewline
\hline 
Component 3 & $0.009$\tabularnewline
\hline 
Component 4 & $-0.047$\tabularnewline
\hline 
\end{tabular}\caption{\textbf{\protect\label{tab:Scalar-curvature-1}Scalar curvature} for
session (L\_RS\_090817) for all four components.}
\end{table}

\begin{table}[H]
\centering
\begin{tabular}{|c|c|}
\hline 
 & Sectional curvature\tabularnewline
\hline 
\hline 
Component 1 & $-0.024$\tabularnewline
\hline 
Component 2 & $-0.048$\tabularnewline
\hline 
Component 3 & $-0.019$\tabularnewline
\hline 
Component 4 & $-0.040$\tabularnewline
\hline 
\end{tabular}\caption{\textbf{\protect\label{tab:Scalar-curvature-2}Scalar curvature }for
session (L\_RS\_100817) for all four components.}
\end{table}

\begin{figure}[H]
\includegraphics{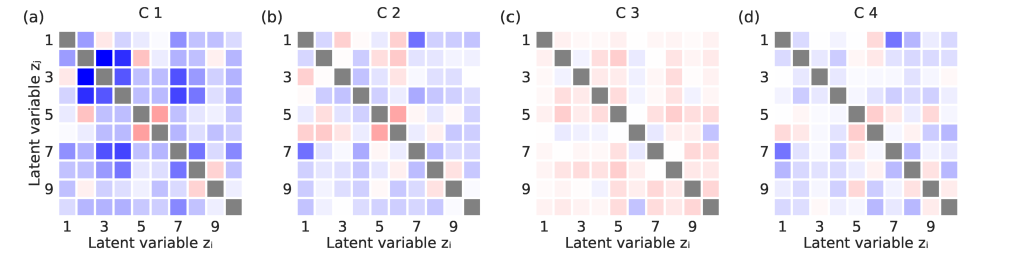}\caption{\protect\label{fig:sectional-curvature-1}\textbf{Geometrical characteristics
of the latent components (session L\_RS\_090817). }Panels\textbf{
(a-d)} show the sectional curvature for each of the four components
of the latent distribution. These values are computed from the quadratically
approximated inverse network. Red indicates positive curvature, while
blue indicates negative curvature. The color scale is consistent across
all panels, with a range defined by the values reported in Table~\ref{tab:Scalar-curvature-1}
for the scalar curvature (i.e.\ the mean of sectional curvatures).}
\end{figure}

\begin{figure}[H]
\includegraphics{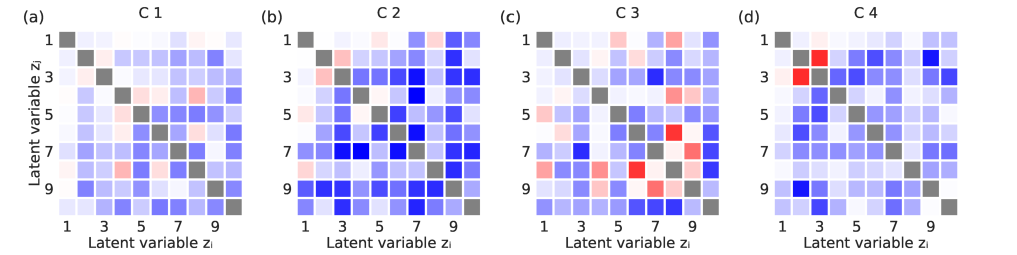}\caption{\protect\label{fig:sectional-curvature-2}\textbf{Geometrical characteristics
of the latent components (session L\_RS\_100817). }Panels\textbf{
(a-d)} show the sectional curvature for each of the four components
of the latent distribution. These values are computed from the quadratically
approximated inverse network. Red indicates positive curvature, while
blue indicates negative curvature. The color scale is consistent across
all panels, with a range defined by the values reported in Table~\ref{tab:Scalar-curvature-2}
for the scalar curvature (i.e.\ the mean of sectional curvatures).}
\end{figure}

\end{widetext}
\end{document}